\documentclass[12pt,a4paper]{article}

\usepackage[T1]{fontenc}
\usepackage[latin1]{inputenc}
\usepackage[nosort]{cite}
\usepackage{color}
\usepackage[bulletsep]{collref}
\usepackage{graphicx}
\usepackage{bbm}
\usepackage{amsmath}
\usepackage{amssymb}
\usepackage{physics}
\usepackage{mathtools}
\usepackage{subfig}
\usepackage{verbatim}
\usepackage{multirow}
\usepackage{tikz}
\usepackage{amsthm}
\usepackage{fancyhdr}
\usepackage{wrapfig}
\usepackage{hyperref}
\usepackage{dsfont}
\usepackage{delimset} 
\usepackage{shuffle} 
\usepackage{mathtools}
\usepackage{enumitem}   

\usepackage{cancel}
\usepackage{subfiles}

\makeatletter \@addtoreset{equation}{section} \makeatother

\makeatletter
\let\old@startsection=\@startsection
\let\oldl@section=\l@section
\renewcommand{\@startsection}[6]{\old@startsection{#1}{#2}{#3}{#4}{#5}{#6\mathversion{bold}}}
\renewcommand{\l@section}[2]{\oldl@section{\mathversion{bold}#1}{#2}}
\makeatother

\makeatletter
\let\old@makecaption=\@makecaption
\def\@makecaption{\small\old@makecaption}
\makeatother

\let\oldPhi=\Phi
\let\oldPsi=\Psi
\let\oldGamma=\Gamma
\let\oldDelta=\Delta
\let\oldSigma=\Sigma
\let\oldTheta=\Theta
\let\oldPi=\Pi
\let\oldUpsilon=\Upsilon
\renewcommand{\Phi}{\mathnormal{\oldPhi}}
\renewcommand{\Psi}{\mathnormal{\oldPsi}}
\renewcommand{\Gamma}{\mathnormal{\oldGamma}}
\renewcommand{\Sigma}{\mathnormal{\oldSigma}}
\renewcommand{\Delta}{\mathnormal{\oldDelta}}
\renewcommand{\Theta}{\mathnormal{\oldTheta}}
\renewcommand{\Pi}{\mathnormal{\oldPi}}
\renewcommand{\Upsilon}{\mathnormal{\oldUpsilon}}

\newcommand{\PP}[3]{\prod_{#1 = #2}^{#3} \left( x_{#1 0}^2 \right)^{a_{#1}}}

\newcommand{\PPE}[2]{\displaystyle \prod_{#1 \in V_{#2}} (x_{#1 #2}^2)^{a_{#1}}}
\newcommand{\PPI}[2]{\displaystyle \prod_{X_{#1} \in V'_{X_{#2}}} (x^2_{X_{#2} X_{#1}})^{b_{#2, #1}}}
\newcommand{\PPIZ}[2]{\displaystyle \prod_{Z_{#1} \in V'_{Z_{#2}}} (x^2_{Z_{#2} Z_{#1}})^{b'_{#2, #1}}}

\newcommand{\superN}{\mathcal{N}}

\newcommand{\gen}[1]{\mathrm{#1}}
\newcommand{\levz}[1]{\mathrm{#1}}
\newcommand{\levo}[1]{ \gen{\widehat #1}}

\newcommand{\Eval}{s} 
\makeatletter
\newlength{\apb@width}
\newcommand{\autoparbox}[2][c]{\settowidth{\apb@width}{#2}\parbox[#1]{\apb@width}{#2}}
\newcommand{\includegraphicsbox}[2][]{\autoparbox{\includegraphics[#1]{#2}}}
\makeatother

\ifx\genfrac\sdflkaj

\else

\fi
\newcommand{\sfrac}[2]{{\textstyle\frac{#1}{#2}}}

\newcommand{\ihalf}{\sfrac{i}{2}}

\makeatletter
\def\mr@ignsp#1 {\ifx\:#1\@empty\else #1\expandafter\mr@ignsp\fi}%
\newcommand{\multiref}[1]{\begingroup
\xdef\mr@no@sparg{\expandafter\mr@ignsp#1 \: }%
\def\mr@comma{}%
\@for\mr@refs:=\mr@no@sparg\do{\mr@comma\def\mr@comma{,}\ref{\mr@refs}}%
\endgroup}
\makeatother

\newcommand{\hypref}[2]{\ifx\href\asklfhas #2\else\href{#1}{#2}\fi}
\newcommand{\Secref}[1]{Section~\multiref{#1}}
\newcommand{\secref}[1]{Section~\multiref{#1}}

\newcommand{\appref}[1]{Appendix~\multiref{#1}}

\newcommand{\Figref}[1]{Figure~\multiref{#1}}

\renewcommand{\eqref}[1]{(\multiref{#1})}

\ifx\href\asklfhas\newcommand{\href}[2]{#2}\fi

\newcommand{\be}{\begin{eqnarray}}
\newcommand{\ee}{\end{eqnarray}}

\begin{document}

\thispagestyle{empty}

\begin{flushright}\footnotesize
\texttt{BONN-TH-2024-14}
\end{flushright}
\vspace{.2cm}

\begin{center}%


{\LARGE\textbf{\mathversion{bold}%
The Feyn-Structure of Yangian Symmetry
}\par}

\vspace{1cm}
{\textsc{Florian Loebbert, Harshad Mathur }}
\vspace{8mm} \\
\textit{%
Bethe Center for Theoretical Physics \\
Universit\"at Bonn, 53115, Germany
}
\vspace{.5cm}

\texttt{loebbert@uni-bonn.de}, \\\texttt{com.mailantiproton@protonmail.com}

\par\vspace{15mm}

\textbf{Abstract} \vspace{5mm}

\begin{minipage}{13cm}
Yangian-type differential operators are shown to constrain Feynman integrals beyond the restriction to integrable graphs.
In particular, we prove that all position-space Feynman diagrams at tree level feature a Yangian level-one momentum symmetry as long as their external coordinates are distinct. This symmetry is traced back to a set of more elementary bilocal operators, which annihilate the integrals. In dual momentum space,
the considered Feynman graphs represent multi-loop integrals without `loops of loops', generalizing for instance the family of so-called train track or train track network diagrams.  The extension of these results to integrals with massive propagators on the boundary of the Feynman graph is established. 
When specializing to the dual conformal case, where propagator powers sum up to the spacetime dimension at each position-space vertex, the symmetry extends to the full dual conformal Yangian. Hence, our findings represent a generalization of the statements on the Yangian symmetry of Feynman integrals beyond integrability and reveal its origin lying in a set of more elementary bilocal annihilators. Previous applications of the Yangian suggest to employ the resulting differential equations for bootstrapping multi-loop integrals beyond the dual conformal case. The considered bilocal constraints on Feynman integrals resemble the definition of conformal partial waves via Casimir operators, but are based on a different algebraic structure.
\end{minipage}
\end{center}

\newpage 

\tableofcontents
\bigskip
\hrule

\section{Introduction and Overview}
Feynman integrals represent a crucial ingredient for calculations within the framework of quantum field theory --- in particular, when it comes to explicit predictions for experiments at particle colliders or gravitational wave observatories. Their understanding is thus of great importance and independent of a specific quantum field theory model. Hence, it is desirable to refine the machinery for the computation of Feynman integrals, such that it exploits all available symmetries, formal insights and computational techniques.
\medskip

One of the most important sources for new formal insights in the last decades has been the Anti-de-Sitter/Conformal Field Theory (AdS/CFT) duality. In particular, in the planar limit integrability implies a large network of mathematical structures that has led to a plethora of new results, relations and techniques for quantum field theory in four spacetime dimensions.
The integrability of the planar AdS/CFT correspondence can be understood to originate from the presence of two conformal symmetries having a natural home on the two sides of the duality. It is an interesting question in how far the mathematical constraints and tools associated with integrability can still be employed for computations in quantum field theory, if one of the two conformal symmetries is given up. It is one goal of the present paper to approach this question in the context of Feynman integrals.
\medskip

A direct connection between the integrability in AdS/CFT and the properties of Feynman integrals is established via the so-called fishnet models \cite{Gurdogan:2015csr,Caetano:2016ydc,Kazakov:2018qbr,Kazakov:2018gcy,Loebbert:2020tje,Kazakov:2022dbd,Alfimov:2023vev}. These are defined by first gamma-deforming the Lagrangian of $\superN=4$ SYM theory, followed by a judicious double-scaling limit in the three gamma-parameters and the original coupling constant. Several of the original fields decouple in this limit, which leads to drastically simplified but non-trivial models of quantum field theory. The simplest example is the bi-scalar fishnet model for which most correlation functions are computed by a single Feynman integral of fishnet structure. This implies that the mathematical properties of these correlation functions are identical to those of individual Feynman integrals. In particular, it can be shown that these Feynman integrals, alias fishnet correlators, feature a Yangian symmetry~\cite{Chicherin:2017cns,Chicherin:2017frs,Loebbert:2019vcj,Loebbert:2020hxk,Loebbert:2022nfu,Kazakov:2023nyu}, which represents the algebraic underpinning of rational integrable models including the planar AdS/CFT correspondence~\cite{Torrielli:2010kq,Torrielli:2011gg,Loebbert:2016cdm}.
\medskip

The conformal Yangian symmetry of Feynman integrals was first shown for square fishnet graphs \cite{Chicherin:2017cns}, which correspond to correlation functions in the above bi-scalar fishnet model. This result was extended to fishnet diagrams cut from all three regular tilings of the plane as well as to brick-wall diagrams with fermionic propagator lines \cite{Chicherin:2017frs}. Moreover, first integrals with parametric propagator powers in generic spacetime dimensions~$D$ were investigated using the Yangian in \cite{Loebbert:2019vcj}, see also the closely related analysis in momentum space \cite{Coriano:2019sth} (cf.\ \Secref{sec:MomSpace}). Yangian symmetry was also found to include diagrams with massive propagators on the boundary \cite{Loebbert:2020hxk}. Finally, the class of the so-called loom diagrams defined via the existence of a dual Baxter lattice was argued to be Yangian invariant in \cite{Kazakov:2023nyu}. Notably, the latter graphs were already associated with integrable vertex models by Zamolodchikov in 1980 \cite{Zamolodchikov:1980mb}.
\medskip

In analogy to the AdS/CFT correspondence, the Yangian symmetry of fishnet Feynman integrals is understood as the closure of a \emph{momentum space conformal symmetry} and a \emph{dual conformal symmetry} \cite{Drummond:2009fd,Chicherin:2017cns,Loebbert:2020glj}. Here, the latter naturally acts on the dual region momenta $x_j^\mu$ defined via the relation $p_j^\mu=x_j^\mu-x_{j+1}^\mu$. The full Yangian is generated by the dual conformal (or level-zero) symmetry and the level-one momentum operator $\levo{P}^\mu$, which will play a distinguished role in the present paper. This operator can be understood as the position space version of a momentum space special conformal symmetry~\cite{Loebbert:2020glj}. 
It is well known that the dual conformal symmetry of Feynman integrals requires that the powers of propagators entering into the same integration vertex, sum up to the spacetime dimension $D$.
In the following, we will refer to this constraint as the \emph{dual conformal} or \emph{level-zero condition}. 
Note that this statement is independent of the momentum space conformal symmetry.
\medskip

In this paper, we address the question of what happens to the above families of Feynman integrals that were identified as being Yangian invariant, if the dual conformal condition is not imposed. That is, on the one hand we want to understand in detail which classes of diagrams feature the non-local Yangian level-one momentum symmetry $\levo{P}^\mu$. On the other hand, the results of \cite{Loebbert:2020glj,Duhr:2022pch,Duhr:2024hjf} suggest that there are more Yangian-type annihilators than those corresponding to  the original generator, which are obtained by applying permutation symmetries of the considered graph to $\levo{P}^\mu$. It is another goal of this paper to properly understand this sub-structure of the full Yangian symmetry.
\medskip

We take the viewpoint that if we understand which integrals have a level-one momentum symmetry, these integrals will also have full Yangian symmetry once the dual conformal condition is imposed on the propagator powers. In that sense, control over the Yangian level-one momentum also implies control over the full Yangian symmetry of Feynman graphs. While translation and Lorentz symmetries are trivially implemented into the considered Feynman graphs, dilatation invariance is achieved by an appropriate choice of the scaling dimensions entering into the representation of the dilatation generator\footnote{Alternatively we can set the scaling dimensions to zero and multiply the integral by an overall function of the external kinematics which compensates the conformal weight.}. The key to understanding Yangian and momentum space conformal symmetry is thus the $x$-space Yangian level-one momentum generator $\levo{P}^\mu$, which corresponds to the momentum space special conformal generator in dual $p$-space, here denoted by~$ \gen{\bar K}^\mu$.
\medskip

A motivation to further extend the above insights on the Yangian symmetry of Feynman graphs to more generic situations comes from the success to bootstrap Feynman integrals using the Yangian.
This idea was first approached in \cite{Loebbert:2019vcj} on the examples of the box, double box and hexagon integrals with massless propagators and for generic Euclidean kinematics.
In \cite{Loebbert:2020hxk,Loebbert:2020glj}, Yangian and momentum space conformal symmetry were  used to bootstrap massive Feynman integrals at one loop. Considering the case of the massless box integral, it was demonstrated that this approach can also be employed in Minkowski kinematics \cite{Corcoran:2020epz}. In \cite{Loebbert:2020aos}, examples of divergent integrals were bootstrapped in dimensional regularization using the Yangian level-one symmetry. Finally, a new connection between the Yangian and Calabi--Yau geometry was established for fishnet integrals in two spacetime dimensions and it was shown that the combination of Yangian and permutation symmetries fixes this geometry completely \cite{Duhr:2022pch,Duhr:2023eld,Duhr:2024hjf}. 
\medskip

\begin{figure}[t]
\begin{center}
\includegraphicsbox[scale=1]{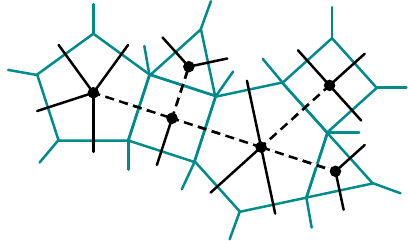}
\end{center}
\caption{Generic position ($x$-)space tree diagram (black) and its dual momentum ($p$-)space graph (green) related via the duality transformation $p_j=x_j-x_{j+1}$. The findings of the present paper require that internal propagators are massless (dashed lines), while external propagators can be either massive or massless (solid lines).}
\label{fig:generaltree}
\end{figure}

These applications in mind, the present work is inspired by the observation in \cite{Loebbert:2020hxk} that the (massive) level-one momentum symmetry persists beyond the integrable (i.e.\ dual conformal) limit for general classes of massless and massive Feynman integrals --- at least at lower loop orders. We will show in this paper that this observation is in fact systematic and generalizes to higher-loop integrals. Moreover, it leads to the discovery of more elementary symmetries that furnish the building blocks of the full Yangian generator. Indications for such a sub-structure of the Yangian symmetry were found in \cite{Loebbert:2020glj,Duhr:2022pch,Duhr:2024hjf}. Our findings are more general than the fully Yangian-symmetric situation, which is obtained when specializing to the dual conformal constraints on the propagator powers. As reviewed in \secref{sec:MomSpace}, the full level-one momentum symmetry can be related to a (massive) momentum space conformal symmetry in dual momentum space. There are, however, reasons why it can be advantageous to work with the non-local $x$-space symmetry: Firstly, in position space the considered Feynman graphs feature a natural set of permutation symmetries that nicely combines with the level-one symmetry, as has been exploited e.g.\ in \cite{Loebbert:2019vcj,Corcoran:2020epz,Duhr:2022pch,Duhr:2024hjf}. Secondly, as we will show in \secref{sec:Bilocal}, the level-one symmetry decomposes into more elementary bilocal annihilators of the integrals, which imply independent constraints and for which an analogue in momentum space seems not obvious.
\medskip

\begin{figure}[t]
\begin{center}
\includegraphicsbox[scale=1]{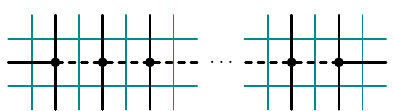}
\qquad\qquad
\includegraphicsbox[scale=1]{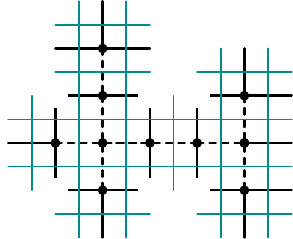}
\end{center}
\caption{Left hand side: Train track diagram in $x$-space (black) and $p$-space (green). Right hand side: Train track network diagram. }
\label{fig:traintrack}
\end{figure}

In particular, we will demonstrate that a natural class of integrals with Yangian level-one symmetry is given by position-space Feynman graphs at tree level with massless internal and either massive or massless external propagators, cf. \Figref{fig:generaltree}.
As we will show explicitly, these feature a Yangian $\levo{P}$-symmetry in full generality, i.e.\ without imposing any constraints on the propagator powers. This is in contrast to their dual conformal symmetry, which only holds if propagator powers at each integration vertex sum up to the spacetime dimension. Moreover, we will demonstrate that this full level-one symmetry decomposes into different classes of bilocal annihilators of the considered diagrams. The considered graphs correspond to challenging multi-loop integrals in the dual momentum space and can be understood as generalizations of the (linear and network) train track integrals studied e.g.\ in \cite{Bourjaily:2018ycu,Vergu:2020uur,Duhr:2022pch,Cao:2023tpx,McLeod:2023doa,Duhr:2024hjf} to generic polygon tracks and networks, see \Figref{fig:generaltree} and \Figref{fig:traintrack}.
Note that here we employ the words integration `loops' or `vertices' interchangeably since they correspond to the same integrals represented in the dual $p$- or $x$-space picture, respectively.
\medskip

Note that the full Yangian symmetry of Feynman integrals was shown using the so-called lasso method \cite{Chicherin:2017cns,Chicherin:2017frs,Kazakov:2023nyu}, which relies on the \emph{RTT-realization} of the Yangian algebra, see e.g.\ \cite{Loebbert:2016cdm}. In this formulation, level-zero and level-one Yangian symmetry are both contained in the monodromy matrix $T(u)$. Expansion in powers of the spectral parameter $u$ leads to the different levels of Yangian generators and Feynman graphs are eigenstates of this monodromy operator. In order to disentangle level-zero and level-one symmetry, it is more useful to employ the \emph{first realization} of the Yangian, which relies on using the explicit level-zero and level-one generators, instead of their generating function $T(u)$. In this formulation, Feynman graphs are annihilated by the level-zero and level-one generators. In particular, this also allows us to understand Feynman integrals with massive propagators generalizing the results of \cite{Loebbert:2020hxk,Loebbert:2020glj}, which so far have not been understood in terms of the RTT-formulation.
\medskip

Throughout this paper, we assume that all considered Feynman graphs have distinct external points. Moreover, we assume the parametric propagator powers to lie in a range which guarantees that the considered integrals are non-singular. We expect, however, that the level-one symmetry also applies to divergent integrals, cf.\ the applications in \cite{Loebbert:2020aos}.

Note that there is a choice in drawing the external legs of a Feynman graph, which relates to our choice of clockwise enumeration of external legs. 
Whenever this choice is relevant for the enumeration of external legs, we assume that all legs attached to internal vertices lie above or left of internal propagators lines. This is convenient since our clockwise numbering of external legs matches the ordering of the integration vertices obtained by going clockwise around the graph. An example is given by the middle vertex of the following three-vertex diagram, where we choose to draw all external legs above the line of internal propagators:
\begin{equation}
\includegraphicsbox[]{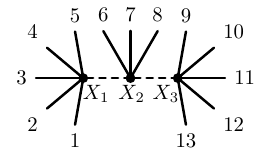}\;,
\qquad \text{with}\qquad
\begin{aligned}
&V_{X_1}=\{1,2,3,4,5\},
\\
&V_{X_2}=\{ 6,7,8\},
\\
&V_{X_3}= \{9,10,11,12,13\}.
\end{aligned}
\end{equation}
When dealing with multiple integration vertices (or loops), it will also be convenient to omit the external legs and to draw e.g.\ the above three-loop graph in the following way:
\begin{equation}
    \includegraphicsbox[]{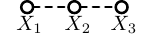}
\end{equation}
Here, we refer to external legs $j$ attached to e.g.\ vertex $X_1$ as $j\in V_{X_1}=\{1,2,3,4,5\}$.

\section{Yangian and Conformal Symmetry}
In this section, we briefly review the extension of the conformal algebra to the infinite dimensional Yangian, as well as the conditions for the conformal symmetry of Feynman graphs in position/region-momentum space ($x$-space).

\subsection{The Yangian} 
Consider the generators of the Euclidean conformal algebra $\mathfrak{so}(1,D+1)$ in $D$ spacetime dimensions:
\begin{equation}
    \levz{J}^a \in \{\levz{P}^{\mu}, \levz{L}^{\mu \nu}, \levz{D}, \levz{K}^{\mu}\}.
\end{equation}
These satisfy the Lie algebra commutation relations
\begin{equation}
    [ \levz{J}^a, \levz{J}^b ] = {f^{ab}}_{c} \, \levz{J}^c,
\end{equation}
where the conformal structure constants ${f^{ab}}_{c}$ can be read off from the individual commutators explicitly given in \appref{app:confalg}. We extend the conformal algebra $\mathfrak{so}(1,D+1)$ to the Yangian $Y[\mathfrak{so}(1,D+1)]$ by adding a set of level-one generators $\levo{J}^a$, which satisfy the commutation relations%
\footnote{This way to define the Yangian corresponds to Drinfeld's original definition called the \emph{first realization}~\cite{Drinfeld:1985rx}.}
\begin{equation}\label{eq:levocomm}
    [ \levz{J}^a, \levo{J}^b ] = {f^{ab}}_{c} \, \levo{J}^c.
\end{equation}
In addition, the level-one generators are constrained by the Serre relations detailed in \appref{app:lev1densities}. Level-zero and level-one generators are sufficient to generate the whole Yangian, which is organized in terms of an infinite number of levels of symmetry generators:
\begin{align}
    &\text{Level zero:} \quad \levz{J}^{(0)}_a = \levz{J}_a,\\
    &\text{Level one:}  \quad\levz{J}^{(1)}_a = \levo{J}_a,\\
    &\text{Level two:}  \quad\levz{J}^{(2)}_a \simeq {f_a}^{bc} \, [ \levo{J}_b, \levo{J}_c ].\\
    &\dots.\nonumber
\end{align}
The above level-zero and level-one generators act on $n$-point Feynman integrals via the following tensor product actions. In particular, the level-zero generators have a trivial tensor product action
\begin{equation}
        \levz{J}^a = \sum^n_{j=1} \levz{J}^a_j, 
        \label{eq:levzcoproduct}
\end{equation}
with the density $\levz{J}^a_j$ acting on external leg $j$. Below, we will specify the explicit representation of these densities in terms of differential operators generating the conformal algebra.
The tensor product action for the level-one generator takes the form
\begin{equation}
    \levo{J}^a = \frac{1}{2} {f^a}_{bc} \sum^{n}_{j=1} \sum^{n}_{k=j+1} \levz{J}^c_j \levz{J}^b_k + \sum_{j=1} \Eval_j \levz{J}^a_j,
\end{equation}
where the constants $\Eval_j$ are called \textit{evaluation parameters} and can be read off from the graph topology, as will be discussed in the subsequent sections. In this paper, an important role is played by the level-one momentum generator, which takes the explicit form
\begin{equation}\label{eq:levoP}
    \levo{P}^{\mu} = \frac{i}{2} \sum^n_{j=1} \sum^n_{k=j+1} \left( \levz{P}^{\mu}_j \levz{D}_k + \levz{P}_{j \nu} \levz{L}^{\mu \nu}_k - \levz{D}_j \levz{P}^{\mu}_k - \levz{L}^{\mu \nu}_j \levz{P}_{k \nu} \right) + \sum^n_{j=1} \Eval_j \levz{P}_j^{\mu}.
\end{equation}
Via \eqref{eq:levocomm}, one can see that the full set of conformal level-zero generators together with this level-one momentum generator is sufficient to generate the entire Yangian.

\subsection{Position Space Conformal Symmetry of Feynman Integrals}
In this section, we briefly illustrate under which conditions position space Feynman graphs obey level-zero conformal symmetry.
Focusing on the level-zero generators with tensor product action \eqref{eq:levzcoproduct}, we will employ the following $x$-space representation of the conformal generator densities:
\begin{equation}\label{eq:levzdensities}
    \begin{aligned}
    \levz{P}^{\mu}_j &= -i \partial^{\mu}_j, 
    &
    \levz{D}_j &= -i( x_{j \mu} \partial^{\mu}_j + \Delta_j ), \\
    \levz{L}^{\mu \nu}_j &= i( x^{\mu}_j \partial^{\nu}_j - x^{\nu}_j \partial^{\mu}_j ), 
    &
    \levz{K}^{\mu}_j &= -i(2 x^{\mu}_j x^{\nu}_j \partial_{j \nu} - x^2_j \partial^{\mu}_j + 2 \Delta_j x^{\mu}_j),
    \end{aligned}
\end{equation}
where $\displaystyle \partial^{\mu}_j := {\partial}/{\partial x_{j \mu}}$ and $\Delta_j$ denotes the conformal weights at site $j$.

\paragraph{One Loop.} 
For illustration, let us first discuss the situation at one loop\footnote{Though working in dual position space, here we refer to the loop in momentum space which is represented by an integration vertex in position space.}. An $n$-point one-loop Feynman integral takes the form
\begin{equation}\label{eq:oneloop}
    I_n^{(1)} = \int \frac{d^D x_0}{\PP{u}{1}{n}},
\end{equation}
where $x_{u0}^\mu := x_u^\mu - x_0^\mu$. Applying the level-zero generators to the above integral, we find
\begin{equation}
    \begin{aligned}
        \levz{P}^{\mu} I_n^{(1)} = 0, &\qquad
        \levz{D} I_n^{(1)} = -i \left[ \sum^n_{i=1} (\Delta_i - 2 a_i) + D \right] I_n^{(1)}, 
        \\
        \levz{L}^{\mu \nu} I_n^{(1)} = 0, &\qquad 
        \levz{K}^{\mu} I_n^{(1)} = -2i \int \frac{d^D x_0}{\PP{j}{1}{n}} \left[ D - \sum^n_{i=1} a_i \right] x^{\mu}_{0} - 2i \sum^n_{i=1} (\Delta_i - a_i) x_{i}^{\mu} I_n^{(1)}.
    \end{aligned}
\end{equation}
It is thus clear that if we specify the conformal weights to take the values of the propagator powers, i.e.\ $\Delta_j = a_j$, and require the conformal constraint $\sum_{j\in \text{vertex}}a_j=D$, all conformal generators annihilate the above one-loop integrals. 
Note that total derivatives under the integral correspond to vanishing boundary terms for an appropriate choice of propagator powers.

\paragraph{General Integral.} 
Extending the above considerations to a general $n$-point Feynman integral $I_n$ in position space, with $\Delta_j=a_j$, we obtain 
\begin{equation}\label{eq:levz-general}
    \begin{aligned}
    \levz{P}^{\mu} I_n = 0, &\qquad
    \levz{D} I_n = -i  \alpha(a_i, b_i) I_n,
     \\
    \levz{L}^{\mu \nu} I_n = 0, &\qquad
    \levz{K}^{\mu} I_n \propto \sum_X \brk[s]3{ D - \sum_{i \in V_X} a_i - \sum_{j \in V'_X} b_j } x^{\mu}_X,
    \end{aligned}
\end{equation}
where $V_X$ and $V'_X$ are the index sets of all external and internal propagators attached to integration vertex $X$, respectively, and we have defined
\begin{equation}\label{eq:alphabb}
    \alpha(a_i, b_i) := \sum_X \brk[s]3{ D - \sum_{i \in V_X} a_i - \sum_{j \in V'_X} b_j }.
\end{equation}
The above shows that the generators $\levz{D}$ and $\levz{K}^{\mu}$ annihilate the considered integrals iff the powers of all propagators (external and internal) attached to a given integration vertex add up to the spacetime dimension $D$:
\begin{equation}
    \sum_{i \in V_X} a_i + \sum_{j \in V'_X} b_j = D \qquad \text{for every integration vertex } X.
\end{equation}
We will refer to this constraint on the propagator powers in a Feynman diagram as the \textit{level-zero} or \emph{dual conformal} condition. Note that in most of the following, we will assume $\Delta_j=a_j$ in the representation of the conformal generators.

\section{Bilocal Symmetries of Feynman Integrals}
\label{sec:Bilocal}

As indicated in the introduction, it is one aim of the present paper to understand the substructure of the Yangian symmetry of Feynman integrals.
In this section, we consider a set of bilocal identities for position space Feynman integrals. On the one hand, these correspond to constraining differential equations on their own. On the other hand, they serve as  building blocks for proving full Yangian level-one momentum symmetry in the subsequent \secref{sec:full}. 

Consider the bilocal level-one momentum generator density defined as
\begin{equation}\label{eq:levobilocal}
    \levo{P}^{\mu}_{jk} := \frac{i}{2} \left( \levz{P}^{\mu}_j \levz{D}_k + \levz{P}_{j \nu} \levz{L}^{\mu \nu}_k - i a_k \levz{P}^{\mu}_j - (j \leftrightarrow k) \right),
\end{equation}
which acts on the external legs $j$ and $k$ only. This operator corresponds to the density under the sum in \eqref{eq:levoP}, with evaluation parameters set to
\begin{equation}
\Eval_j \to +\frac{a_k}{2} ,
\qquad
\Eval_k\to -\frac{a_j}{2} ,
\end{equation}
which is motivated by the below action on a product of propagators \eqref{eq:twopointsym}.
Plugging in the level-zero densities \eqref{eq:levzdensities}, the above operator takes the explicit form
\begin{equation}
    \levo{P}^{\mu}_{jk} = \frac{i}{2} \left( x_{jk \nu} \partial_j^{\mu} \partial_k^{\nu} - x_{jk}^{\mu} \partial_j^{\nu} \partial_{k \nu} + x_{jk \nu} \partial_{j}^{\nu} \partial_k^{\mu} - 2 a_k \partial_j^{\mu} + 2 a_j \partial_k^{\mu} \right).
\end{equation}
For completeness, we list the level-one densities associated with the other conformal generators in \appref{app:lev1densities}.
Applying this second order differential operator to a generic product of two propagators, we find the important identity
\begin{equation}
    \levo{P}^{\mu}_{jk} \left( \frac{1}{(x^2_{j X})^{a_j} (x^2_{k Y})^{a_k}} \right) = 2 i a_j a_k \frac{T^{\alpha \beta \mu \nu} x_{X Y \nu} \ x_{j X \alpha} \ x_{k Y \beta}}{(x^2_{j X})^{a_j + 1} (x^2_{k Y})^{a_k + 1}},
    \label{eq:Phattwoprop}
\end{equation}
where we have defined
\begin{equation}
    T^{\alpha \beta \mu \nu} := \eta^{\alpha \mu} \eta^{\beta \nu} + \eta^{\alpha \nu} \eta^{\beta \mu} - \eta^{\alpha \beta} \eta^{\mu \nu},
\end{equation} 
and $X$ and $Y$ represent arbitrary internal coordinates of the considered graph.

\subsection{Two-Point Symmetries}
\begin{figure}[t]
\begin{center}
\includegraphicsbox{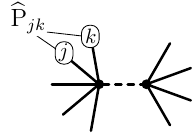} \quad $=0$
\caption{Example of a two-point symmetry annihilating a two-loop graph. All two-point level-one momentum densities $\levo{P}_{jk}^\mu$ with legs $j$ and $k$ attached to the same integration vertex provide annihilators of the integral.}
\label{fig:twopointsym}
\end{center}
\end{figure}

In particular, the above equation \eqref{eq:Phattwoprop} implies that if the level-one momentum generator density is applied to two external legs attached to the same integration vertex, i.e.\ $X = Y$, one gets
\begin{equation}
    \levo{P}^{\mu}_{jk} \left( \frac{1}{(x^2_{j X})^{a_j} (x^2_{k X})^{a_k}} \right) = 0.
    \label{eq:twopointsym}
\end{equation}
Note that this invariance implies a set of second order differential equations for generic Feynman integrals for all pairs of legs attached to the same vertex. We will refer to these as \emph{two-point symmetries}, which can be applied to any graph with two legs attached to the same vertex cf.\ \Figref{fig:twopointsym}. Similarly, one can show that the other two-point level-one generator densities defined in \appref{app:lev1densities} annihilate the above product of propagators.

\subsection{Two-Vertex Symmetries}
\label{sec:simple}
In this subsection, we discuss several instances of bilocal symmetries, where each of the two legs of the level-one momentum generator acts on all legs of (at least) two different integration vertices.

\subsubsection{End-Vertex Symmetries}
Consider an arbitrary $n$-point diagram that contains an integration vertex $X$ (henceforth referred to as end-vertex) that is only attached to one other integration vertex $Y$ by an internal propagator with power $b$. The general form of the corresponding Feynman integral is given by
\begin{equation}
    I_n = \int \cdots \frac{d^Dx_X d^Dx_Y}{(x^2_{XY})^b \PPE{u}{X} \PPE{v}{Y}} \cdots
    =
    \includegraphicsbox[]{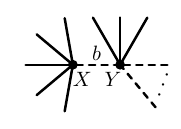}
\end{equation}

\noindent We now define the following operators
\begin{align}
    \levz{P}^{\mu}_{X} &:= \sum_{j \in V_X} \levz{P}^{\mu}_{j}, \label{eq:levzP-partial} \\
    \levo{P}^{\mu}_{XY} &:= \sum_{j \in V_X} \sum_{k \in V_Y} \levo{P}^{\mu}_{jk}, \label{eq:levoP-partial}
\end{align}
which will play a significant role in the subsequent calculations. Recall that the action of the bilocal level-one momentum generator density on the product of propagators \eqref{eq:Phattwoprop} can be written as
\begin{equation}
    \levo{P}^{\mu}_{jk} \left( \frac{1}{(x^2_{j X})^{a_j} (x^2_{k Y})^{a_k}} \right) = \frac{i}{2} T^{\alpha \beta \mu \nu} x_{X Y \nu} \ \frac{2 a_j x_{j X \alpha}}{(x^2_{j X})^{a_j + 1}} \, \frac{2 a_k x_{k Y \beta}}{(x^2_{k Y})^{a_k + 1}},
\end{equation}
and thus, the action of the operator \eqref{eq:levoP-partial} on the integral takes the form%
\footnote{This is a proportionality on the level of the integrand.}
\begin{equation}
    \levo{P}^{\mu}_{XY} I_n \propto \frac{i}{2} T^{\alpha \beta \mu \nu} \frac{x_{X Y \nu}}{(x^2_{XY})^b} \sum_{j \in V_X} \partial_{j \alpha} \frac{1}{\PPE{u}{X}} \, \sum_{k \in V_Y} \partial_{k \beta} \frac{1}{\PPE{v}{Y}}.
\end{equation}
Now, we can re-express the $\partial_{k\beta}$-derivative as
\begin{equation}
    i \sum_{k \in V_Y} \partial_{k \beta} \frac{1}{\PPE{v}{Y}} \cdots = -\levz{P}_{Y \beta} \frac{1}{\PPE{v}{Y}} \cdots,
\end{equation}
while the $\partial_{j\alpha}$-derivative turns into
\begin{equation}
        \sum_{j \in V_X} \partial_{j \alpha} \frac{1}{\PPE{u}{X}} = -\partial_{X \alpha} \frac{1}{\PPE{u}{X}}, 
\end{equation}
such that
\begin{equation}
        -\frac{x_{X Y \nu}}{(x^2_{XY})^b} \, \partial_{X \alpha} \frac{1}{\PPE{u}{X}} = \frac{1}{\PPE{u}{X}} \partial_{X \alpha} \frac{x_{X Y \nu}}{(x^2_{XY})^b} + \text{total derivative},
\end{equation}
with the total derivative with respect to $X$ vanishing upon integration. Evaluating the derivative and recombining the different terms, we get
\begin{equation}
    \levo{P}^{\mu}_{XY} I_n \propto -\frac{1}{2} \levz{P}_{Y \beta} \left[ T^{\alpha \beta \mu \nu} \left( \eta_{\alpha \nu} - \frac{2 b \, x_{XY \nu} \, x_{XY \alpha}}{x^2_{XY}} \right) \cdots \right].
\end{equation}
Completing the contraction with $T^{\alpha \beta \mu \nu}$ finally gives
\begin{equation}\label{eq:endvertex}
    \levo{P}^{\mu}_{XY} I_n = -\frac{1}{2} \left( D - 2b \right) \levz{P}^{\mu}_{Y} I_n.
\end{equation}
In conclusion, we thus find the symmetry equation
\begin{equation}
    \brk[s]2{\levo{P}^{\mu}_{XY} + \frac{1}{2} \left( D - 2b \right) \levz{P}^{\mu}_Y} I_n = 0.
\end{equation}
This equation illustrates the general principle further employed in the subsequent sections: we show that the action of the bilocal operator $\levo{P}$ reduces to the action of local operators $\levz{P}_j$, whose contributions can thus be understood as contributions to the evaluation parameters entering into the level-one momentum generator \eqref{eq:levoP}.
It will be useful to compactify our graphical notation, e.g.\ 
\begin{equation}
\includegraphicsbox[scale=1]{FigEndVertexGraphLegs.pdf}
\quad\to\quad
\includegraphicsbox[scale=1]{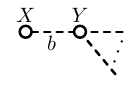}.
\end{equation}
Here, on the right hand side\footnote{We stress that these are not vacuum diagrams.} we omit external legs
and a small circle denotes an integration vertex.
Employing this simple pictorial language, the above two-vertex symmetry is illustrated as
\begin{equation}
\includegraphicsbox[scale=1]{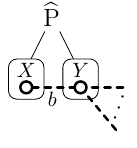}=0.
\end{equation}

\subsubsection{Bridge-Vertex Symmetries}
Now, consider an arbitrary $n$-point diagram that contains two end-vertices, $X$ and $Z$, that are attached to the same integration vertex $Y$ by internal propagators with powers $b$ and $b'$, respectively. The corresponding Feynman integral thus takes the form
\begin{equation}
    I_n = \int \cdots \frac{d^Dx_{X} d^Dx_{Z} d^Dx_Y}{(x^2_{X Y})^{b} (x^2_{YZ})^{b'} \PPE{u}{X} \PPE{v}{Y} \PPE{w}{Z}} \cdots.
\end{equation}
This setup can be illustrated as
\begin{equation}
    I_n = 
    \includegraphicsbox{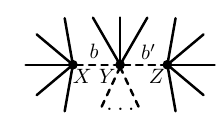}
    \equiv
    \includegraphicsbox{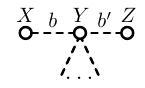}
\end{equation}
Due to the above the end-vertex symmetries, we know that $\levo{P}^{\mu}_{XY}$ and $\levo{P}^{\mu}_{YZ}$ can be reduced to first order differential operators, when acting on the integral. We now act with $\levo{P}^{\mu}_{XZ}$ on the integral:
\begin{equation}
    \levo{P}^{\mu}_{XZ} I_n \propto \frac{i}{2} T^{\alpha \beta \mu \nu} \frac{x_{XY \nu} + x_{YZ \nu}}{(x^2_{XY})^{b} (x^2_{YZ})^{b'}} \sum_{j \in V_X} \partial_{j \alpha} \frac{1}{\PPE{u}{X}} \, \sum_{k \in V_Z} \partial_{k \beta} \frac{1}{\PPE{w}{Z}},
\end{equation}
where $x_{XZ \nu}$ appearing in the action has been rewritten as
\begin{equation}
    x_{XZ \nu} = x_{XY \nu} + x_{YZ \nu}.
\end{equation}
Considering the term containing $x_{X Y \nu}$, the derivative summed over $V_Z$ here is converted into the momentum generator density, whereas the derivative summed over $V_X$ is converted into a derivative with respect to $X$ that can be integrated by parts. For the term with $x_{Y Z \nu}$, the opposite procedure is adopted. This gives us
\begin{equation}
    \begin{aligned}
        i\frac{x_{XY \nu}}{(x^2_{XY})^{b}} \sum_{j \in V_X} \partial_{j \alpha} \left( \cdots \right) \sum_{k \in V_Z} \partial_{k \beta} \left( \cdots \right) &\to -\levz{P}_{Z \beta} \left[ \partial_{X \alpha} \left( \frac{x_{XY \nu}}{(x^2_{XY})^{b}} \right) \cdots \right], \\
        i\frac{x_{YZ \nu}}{(x^2_{YZ})^{b'}} \sum_{j \in V_X} \partial_{j \alpha} \left( \cdots \right) \sum_{k \in V_Z} \partial_{k \beta} \left( \cdots \right) &\to -\levz{P}_{X \alpha} \left[ \partial_{Z \beta} \left( \frac{x_{YZ \nu}}{(x^2_{YZ})^{b'}} \right) \cdots \right].
    \end{aligned}
\end{equation}
Computing the derivatives and contracting the results with $T^{\alpha \beta \mu \nu}$ yields
\begin{equation}
    \begin{aligned}
        T^{\alpha \beta \mu \nu} \partial_{X \alpha} \left( \frac{x_{X Y \nu}}{(x^2_{XY})^{b}} \right) &= \frac{T^{\alpha \beta \mu \nu}}{(x^2_{XY})^{b}} \left( \eta_{\alpha \nu} - \frac{2b \, x_{XY \nu} \, x_{XY \alpha}}{x^2_{XY}} \right) = \eta^{\beta \mu} \frac{D - 2b}{(x^2_{XY})^{b}}, \\
        T^{\alpha \beta \mu \nu} \partial_{Z \beta} \left( \frac{x_{Y Z \nu}}{(x^2_{YZ})^{b'}} \right) &= \frac{T^{\alpha \beta \mu \nu}}{(x^2_{YZ})^{b'}} \left( -\eta_{\beta \nu} + \frac{2b' \, x_{YZ \nu} \, x_{YZ \beta}}{x^2_{YZ}} \right) = -\eta^{\alpha \mu} \frac{D - 2b'}{(x^2_{YZ})^{b'}},
    \end{aligned}
\end{equation}
and upon reinserting the results, we finally get
\begin{equation}\label{eq:bridgevertex}
    \levo{P}^{\mu}_{XZ} I_n = \frac{1}{2} \left( D - 2b' \right) \levz{P}^{\mu}_{X} I_n - \frac{1}{2} \left( D - 2b \right) \levz{P}^{\mu}_{Z} I_n.
\end{equation}
This implies that the operator
\begin{equation}
    \levo{P}^{\mu}_{XZ} - \frac{1}{2} \left( D - 2b' \right) \levz{P}^{\mu}_X + \frac{1}{2} \left( D - 2b \right) \levz{P}^{\mu}_Z,
\end{equation}
annihilates the integral and is thus a symmetry generator. This symmetry is graphically represented as
\begin{equation}
\includegraphicsbox[]{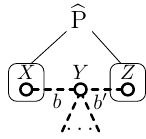}=0.
\end{equation}

\subsection{Multi-Vertex Symmetries}
\label{sec:general}
In this subsection, we generalize the above symmetries to situations where at least one of the two legs of the bilocal level-one generator acts on a sum of integration vertices.

\subsubsection{Special End-Vertex Symmetries}
Let us first discuss the special case of an invariance that follows from purely algebraic considerations and the properties under level-zero conformal generators, cf.\ \cite{Loebbert:2020glj}.
Consider an arbitrary $n$-point Feynman diagram with vertices $\{ X_1, X_2, \dots \}$. We want to study the following bilocal operator associated to the level-one momentum:
\begin{equation}
    \sfrac{1}{2} {f^{\levz{P}^{\mu}}}_{bc} \sum_{1<q} \levz{J}^c_{X_1} \levz{J}^b_{X_q} = \sfrac{1}{2} {f^{\levz{P}^{\mu}}}_{bc} \levz{J}^c_{X_1} \left( \levz{J}^b - \levz{J}^b_{X_1} \right) = \ihalf \left( \levz{P}^{\mu}_{X_1} \levz{D} + \levz{P}_{X_1 \nu} \levz{L}^{\mu \nu} - \levz{D}_{X_1} \levz{P}^{\mu} - \levz{L}^{\mu \nu}_{X_1} \levz{P}_{\nu} \right) - \frac{\mathfrak{c}}{2} \levz{P}^{\mu}_{X_1}.
\end{equation}
Here, we used the identity
\begin{equation}
    \frac{1}{2} {f^{\levz{P}^{\mu}}}_{bc} \levz{J}^c_{X_1} \levz{J}^b_{X_1}
    = \frac{1}{4} {f^{\levz{P}^{\mu}}}_{bc} \left[ \levz{J}^c_{X_1}, \levz{J}^b_{X_1} \right]
    = \frac{1}{4} {f^{\levz{P}^{\mu}}}_{bc} {f^{cb}}_d \levz{J}^d_{X_1}
    = \frac{\mathfrak{c}}{2} \levz{P}^{\mu}_{X_1},
\end{equation}
where $\mathfrak{c}$ denotes the dual Coxeter number, which we define by the relation ${f^a}_{bc} {f^{cb}}_d = 2 \mathfrak{c} \delta^a_d$ and whose value can be computed as
\begin{equation}
    \mathfrak{c} \levz{P}^{\mu}
    = \frac{1}{2} {f^{\levz{P}^{\mu}}}_{bc} \left[ \levz{J}^c, \levz{J}^b \right]
    = \frac{1}{2} \left( i \left[ \levz{P}^{\mu}, \levz{D} \right] - i \left[ \levz{D}, \levz{P}^{\mu} \right] + i \left[ \levz{P}_{\nu}, \levz{L}^{\mu \nu} \right] - i \left[ \levz{L}^{\mu \nu}, \levz{P}_{\nu} \right] \right)
    = D \levz{P}^{\mu}.
\end{equation}
Thus, the dual Coxeter number $\mathfrak{c}$ equals the spacetime dimension $D$. Using \eqref{eq:levz-general}, the action of the above operator finally yields
\begin{equation}
\label{eq:multivertex}
   \brk[s]3{\frac{1}{2} {f^{\levz{P}^{\mu}}}_{bc} \sum_{1<q} \levz{J}^c_{X_1} \levz{J}^b_{X_q} - \frac{1}{2} \left( \alpha(a_i, b_i) - D \right) \levz{P}^{\mu}_{X_1}} I_n = 0,
\end{equation}
where we remind of the above abbreviation \eqref{eq:alphabb}
\begin{equation}
\alpha(a_i, b_i)=\sum_X \brk[s]3{ D - \sum_{i \in V_X} a_i - \sum_{j \in V'_X} b_j }.
\end{equation}

\subsubsection{Generalized End-Vertex Symmetries}
The simple end-vertex symmetries discussed in \Secref{sec:simple} can be generalized to situations where the end-vertex is replaced by a tree (sub-)graph of the full Feynman integral under consideration, i.e.\ to an invariance of the form
\begin{equation}
    \includegraphicsbox{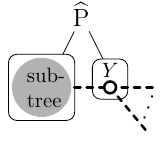}=0.
\end{equation}
Here, the sub-tree consisting of $N$ vertices $\{ X_1, X_2, \cdots, X_{N} \}$ is connected to the rest of the graph via the vertex $Y$. Also for this situation, one finds that the operator
\begin{equation}
    \sum^{N}_{p=1} \levo{P}^{\mu}_{X_{p} Y},
\end{equation}
can be reduced to a first-order differential operator, when acting on the Feynman integral. To prove this, we start by checking the action of $\levo{P}^{\mu}_{X_{p} Y}$ on the integral:
\begin{equation}
    \levo{P}^{\mu}_{X_p Y} I_n \propto \frac{i}{2} T^{\alpha \beta \mu \nu} \frac{x_{X_p Y \nu}}{\PPI{r}{p}} \sum_{j \in V_{X_p}} \partial_{j \alpha} \frac{1}{\PPE{u}{X_p}} \, \sum_{k \in V_Y} \partial_{k \beta} \frac{1}{\PPE{v}{Y}}.
\end{equation}
Here, $V'_{X_p}$ refers to the set of internal propagators attached to $X_p$.
The $\partial_{k \beta}$-derivative is written as
\begin{equation}
    i \sum_{k \in V_Y} \partial_{k \beta} \frac{1}{\PPE{v}{Y}} \cdots = -\levz{P}_{Y \beta} \frac{1}{\PPE{v}{Y}} \cdots,
\end{equation}
while the $\partial_{j\alpha}$-derivative transforms into
\begin{equation}
        \sum_{j \in V_{X_p}} \partial_{j \alpha} \frac{1}{\PPE{u}{X_p}} = -\partial_{X_p \alpha} \frac{1}{\PPE{u}{X_p}}, 
\end{equation}
such that
\begin{equation}
        -\frac{x_{X_p Y \nu}}{\PPI{r}{p}} \, \partial_{X_p \alpha} \frac{1}{\PPE{u}{X_p}} = \frac{1}{\PPE{u}{X_p}} \partial_{X_p \alpha} \frac{x_{X_p Y \nu}}{\PPI{r}{p}} + \text{tot.~deriv.}
\end{equation}
Once again, the total derivative with respect to $X_p$ vanishes under the integral. Evaluating the resulting derivative and reinserting the transformed quantities, we get
\begin{equation}\label{eq:endvertex-gen-partialint}
    \levo{P}^{\mu}_{X_p Y} I_n \propto -\frac{1}{2} \levz{P}_{Y \beta} \left[ T^{\alpha \beta \mu \nu} \left( \eta_{\alpha \nu} - \sum_{X_r \in V'_{X_p}} \frac{2 b_{p,r} \, x_{X_p Y \nu} \, x_{X_{p} X_{r} \alpha}}{x^2_{X_{p} X_{r}}} \right) \cdots \right].
\end{equation}
Now, consider $X_{p'} \in V'_{X_p}$ i.e.\ $X_p$ and $X_{p'}$ are adjacent vertices connected by an internal propagator of power $b_{p,p'}$:
\begin{equation}
    \includegraphicsbox{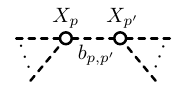}
\end{equation}
The non-trivial contribution from this configuration to the above expression \eqref{eq:endvertex-gen-partialint} reads
\begin{equation}
    -\frac{1}{2} \levz{P}_{Y \beta} \left[ T^{\alpha \beta \mu \nu} \frac{2 b_{p,p'} \, x_{X_{p} Y \nu} \, x_{X_{p} X_{p'} \alpha}}{x^2_{X_p X_{p'}}} \cdots \right].
\end{equation}
Similarly, when acting with $\levo{P}^{\mu}_{X_{p'} Y}$ on the integral, one can extract
\begin{equation}
    -\frac{1}{2} \levz{P}_{Y \beta} \left[ T^{\alpha \beta \mu \nu} \frac{2 b_{p,p'} \, x_{X_{p'} Y \nu} \, x_{X_{p'} X_p \alpha}}{x^2_{X_{p'} X_p}} \cdots \right].
\end{equation}
Thus, from the action of $\levo{P}^{\mu}_{X_{p} Y} + \levo{P}^{\mu}_{X_{p'} Y}$ on the integral, one gets
\begin{equation}
    T^{\alpha \beta \mu \nu} \frac{2 b_{p, p'} \, (x_{X_{p} Y \nu} - x_{X_{p'} Y \nu}) \, x_{X_p X_{p'} \alpha}}{x^2_{X_p X_{p'}}} = 2 b_{p,p'} \, \eta^{\beta \mu}.
\end{equation}
This procedure is performed for every pair of adjacent vertices in the sub-trees. The only non-trivial term remaining at the end comes from the interaction between the vertices $Y$ and $X_1$, and can directly be simplified as
\begin{equation}
    T^{\alpha \beta \mu \nu} \frac{2b_1 \, x_{X_1 Y \nu} \, x_{X_1 Y \alpha}}{x^2_{X_1 Y}} = 2 b_1 \, \eta^{\beta \mu},
\end{equation}
were $b_1\equiv b_{1,Y}$.
After contracting all the terms with $T^{\alpha \beta \mu \nu}$, we finally obtain
\begin{equation}\label{eq:endvertex-gen}
    \sum^{N}_{p=1} \levo{P}^{\mu}_{X_{p} Y} I_n = -\frac{1}{2} \left( N D - 2 \sum^{N}_{i=1} b_i \right) \levz{P}^{\mu}_{Y} I_n.
\end{equation}
Here, $b_i$ corresponds to the power of the propagator before vertex $X_i$.
One can see that if the sub-tree consists of a single vertex, then the above expression reduces to \eqref{eq:endvertex}.


\subsubsection{Generalized Bridge-Vertex Symmetries}
Similarly, the simple bridge-vertex symmetries discussed in \Secref{sec:simple} can be generalized to situations where both vertices connected to a bridge-vertex are replaced by generic tree-level sub-graphs. We will now show an invariance of the schematic form
\begin{equation}
    \includegraphicsbox{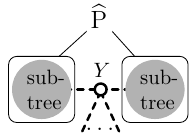}=0\;,
\end{equation}
where the sub-trees with $N_1$ vertices $\{ X_1, X_2, \cdots, X_{N_1} \}$ and $N_2$ vertices $\{ Z_1, Z_2, \cdots, Z_{N_2} \}$ are connected to the rest of the graph and each other via the bridge-vertex $Y$. Here, the operator
\begin{equation}
    \sum^{N_1}_{p=1} \sum^{N_2}_{q=1} \levo{P}^{\mu}_{X_{p} Z_{q}},
\end{equation}
gives rise to a symmetry equation. We begin the proof by acting with $\levo{P}^{\mu}_{X_{p} Z_{q}}$ on the integral:
\begin{equation}
    \levo{P}^{\mu}_{X_p Z_q} I_n \propto \frac{i}{2} T^{\alpha \beta \mu \nu} \frac{x_{X_p Y \nu} + x_{Y Z_q \nu}}{\PPI{r}{p} \PPIZ{s}{q}} \sum_{j \in V_{X_p}} \partial_{j \alpha} \left( \cdots \right) \, \sum_{k \in V_{Z_q}} \partial_{k \beta} \left( \cdots \right),
\end{equation}
where $x_{X_p Z_q \nu}$ was split according to
\begin{equation}
    x_{X_p Z_q \nu} = x_{X_p Y \nu} + x_{Y Z_q \nu}.
\end{equation}
Just like for the simple bridge-vertex symmetry, we transform the two terms in the opposite manner: for the $x_{X_p Y \nu}$-term, the sum of the derivatives over $V_{Z_q}$ becomes the sum of the momentum generator densities over $V_{Z_q}$, while the sum of the derivatives over $V_{X_p}$ becomes a derivative with respect to $X_p$, which is then partially integrated (and vice versa for the $x_{Y Z_q \nu}$-term). This results in
\begin{equation}
    \levo{P}^{\mu}_{X_p Z_q} I_n \propto -\frac{1}{2} \levz{P}_{Z_q \beta} \brk[s]4{ \partial_{X_p \alpha} \brk4{ \frac{T^{\alpha \beta \mu \nu} x_{X_p Y \nu}}{\PPI{r}{p}} } \cdots } - \frac{1}{2} \levz{P}_{X_p \alpha} \brk[s]4{ \partial_{Z_q \beta} \brk4{ \frac{T^{\alpha \beta \mu \nu} x_{Y Z_q \nu}}{\PPIZ{s}{q}} } \cdots }.
\end{equation}
In the first term, the derivative with respect to $X_p$ expands to 
\begin{equation}
    -\frac{1}{2} \levz{P}_{Z_q \beta} \left[ T^{\alpha \beta \mu \nu} \left( \eta_{\alpha \nu} - \sum_{X_r \in V'_{X_p}} \frac{2 b_{p,r} \, x_{X_p Y \nu} \, x_{X_{p} X_{r} \alpha}}{x^2_{X_{p} X_{r}}} \right) \cdots \right],
\end{equation}
and in the second term, the derivative with respect to $Z_q$ gives
\begin{equation}
    -\frac{1}{2} \levz{P}_{X_p \alpha} \left[ T^{\alpha \beta \mu \nu} \left( -\eta_{\beta \nu} + \sum_{Z_r \in V'_{Z_q}} \frac{2 b'_{q,r} \, x_{Z_q Y \nu} \, x_{Z_{q} Z_{r} \beta}}{x^2_{Z_{q} Z_{r}}} \right) \cdots \right].
\end{equation}
Now, for fixed $Z_q$ in the first expression and $X_p$ in the second expression, we repeat the procedure outlined in the above section discussing the generalized end-vertex symmetry, which yields
\begin{equation}
    -\frac{1}{2} \left( N_1 D - 2 \sum^{N_1}_{i=1} b_i \right) \levz{P}^{\mu}_{Z_q}, \qquad \frac{1}{2} \left( N_2 D - 2 \sum^{N_2}_{i=1} b'_i \right) \levz{P}^{\mu}_{X_p},
\end{equation}
respectively. Finally, after adding up all terms, we get
\begin{equation}\label{eq:bridgevertex-gen}
    \sum^{N_1}_{p=1} \sum^{N_2}_{q=1} \levo{P}^{\mu}_{X_{p} Z_{q}} I_n = \frac{1}{2} \left( N_2 D - 2 \sum^{N_2}_{i=1} b'_i \right) \sum^{N_1}_{p=1} \levz{P}^{\mu}_{X_p} I_n - \frac{1}{2} \left( N_1 D - 2 \sum^{N_1}_{i=1} b_i \right) \sum^{N_2}_{q=1} \levz{P}^{\mu}_{Z_q} I_n.
\end{equation}
Again, one can see that if the sub-trees were just single vertices, then the result boils down to \eqref{eq:bridgevertex}.

In the following section, we will demonstrate how the bilocal symmetries discussed in this section can be combined to the full Yangian level-one momentum symmetry of Feynman integrals.

\section{Full $\levo{P}$-Symmetry of Feynman Integrals}
\label{sec:full}
Based on the above bilocal annihilators, in this section, we will discuss which position space Feynman integrals are invariant under the \emph{full} level-one momentum symmetry $\levo{P}$, which acts on all external legs of a given graph and is most natural from the viewpoint of integrability. We will not assume conformal level-zero symmetry, i.e.\ the external and internal propagator powers $a_j$ and $b_j$, respectively, represent unconstrained parameters. Hence, in contrast to the proofs of Yangian symmetry given in \cite{Chicherin:2017cns,Chicherin:2017frs,Kazakov:2023nyu}, which employ the monodromy matrix as the generating function incorporating all Yangian generators, we will follow the approach to investigate the Yangian by understanding the action of the level-one momentum generator on a given graph. 

\subsection{Evaluation Parameters}
\label{sec:Eval}
\begin{figure}[t]
\begin{center}
\includegraphicsbox{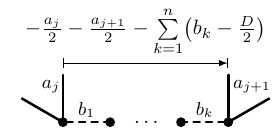}
\end{center}
\caption{Part of the boundary of a position space Feynman graph. The evaluation parameters $\Eval_j$ entering into the Yangian level-one generator are defined recursively. In particular, $\Eval_{j+1}$ is obtained from $\Eval_j$ by adding the terms on top of the arrow connecting the two subsequent legs. Here, we distinguish external propagators (solid lines) and internal propagators (dashed lines), which connect the two neighboring external points $j$ and $j+1$.}
\label{fig:evalrule}
\end{figure}

In this setup, Yangian level-one momentum invariance means that there is a choice of evaluation parameters such that $\levo{P}^\mu$, as given in \eqref{eq:levoP}, annihilates the graph.
In particular, we will prove that also in the absence of level-zero symmetry, invariance under the full level-one momentum requires the evaluation parameters to be chosen according to the following rule, which agrees with the rule identified in the context of the full Yangian symmetry \cite{Chicherin:2017cns,Loebbert:2020hxk,Kazakov:2023nyu}:
\begin{itemize}
\item Choose one of the external legs to be leg 1 and label the other legs in a clockwise fashion.
\item Set $\Eval_1$ to some value of choice. This choice corresponds to the freedom to shift the level-one momentum generator $\levo{P}^\mu$ by a full level-zero momentum generator $\levz{P}^\mu$, which annihilates the graph.
\item The rest of the evaluation parameters is fixed recursively according to, cf.\ \Figref{fig:evalrule}:
\begin{equation}
   \Eval_{j+1}=\Eval_j-\frac{a_j}{2}-\frac{a_{j+1}}{2}-\sum\limits_{k=1}^{n} \brk*{b_{k}-\frac{D}{2}}.
\end{equation}
\end{itemize}
In \secref{sec:alltrees}, we will prove that this general prescription follows from the bilocal symmetries discussed in the previous \Secref{sec:Bilocal}.


\subsection{One-, Two- and Three-Loop Integrals}
In this subsection, we demonstrate that the invariance of tree integrals with one, two and three integration vertices is a simple corollary of the above bilocal annihilators discussed in \Secref{sec:Bilocal}. Similar proofs at one- and two-loop orders were already given in \cite{Loebbert:2020hxk,Loebbert:2020glj}.

\paragraph{One Loop.}
Consider once again the $n$-point one-loop Feynman integral with generic propagator powers:
\begin{equation}
    I_n^{(1)} = \int \frac{d^D x_0}{\PP{u}{1}{n}}.
\end{equation}
In order to show the invariance of the above integral under the action of the level-one momentum generator, we will first re-express the operator in terms of quantities familiar from the above discussions:
\begin{equation}
    \levo{P}^{\mu} = \sum^{n}_{j=1} \sum^{n}_{k=j+1} \levo{P}^{\mu}_{jk} - \frac{1}{2} \sum^n_{j=1} \sum^n_{k=j+1} \left( a_k \levz{P}^{\mu}_j - a_j \levz{P}^{\mu}_k \right) + \sum^n_{j=1} \Eval^{(1)}_j \levz{P}_j^{\mu},
\end{equation}
with the bilocal generator density being defined in \eqref{eq:levobilocal}. Using the identity
\begin{equation}\label{eq:doublecounting}
    \sum^n_{p=1} \sum^n_{q=p+1} a_p b_q = \sum^n_{p=1} \sum^{p-1}_{q=1} a_q b_p,
\end{equation}
and the two-point symmetries \eqref{eq:twopointsym}, we get
\begin{equation}
    \levo{P}^{\mu} I_n^{(1)} = \sum^n_{j=1} \left[ \Eval^{(1)}_j - \frac{1}{2} \sum^n_{k=j+1} a_k + \frac{1}{2} \sum_{k=1}^{j-1} a_j \right] \levz{P}^{\mu}_j I_n^{(1)}.
\end{equation}
Thus, invariance under the level-one momentum generator requires the one-loop evaluation parameter to be defined as
\begin{equation}
\label{eq:eval1loop}
    s^{(1)}_j := \frac{1}{2} \left( \sum_{k=j+1}^n a_k - \sum_{k=1}^{j-1} a_k \right).
\end{equation}
 Similarly, using the above evaluation parameters, the remaining level-one generators can be expressed as double sums of the two-point level-one generator densities specified in \appref{app:lev1densities}, which annihilate the $n$-point one-loop Feynman integral. 

Note that it can be shown that the set of all two-point symmetries $\levo{P}_{jk}^\mu$ is equivalent to full level-one invariance under $\levo{P}^\mu$ combined with the permutation invariance of a generic one-loop graph acting on its external legs, see \cite{Loebbert:2020glj}.

\paragraph{Two Loops.} 
Let us now move on to the Feynman graph with two integration vertices: 
\begin{equation}
    I^{(2)}_n = \int \frac{d^Dx_{X_1} d^Dx_{X_2}}{(x^2_{X_1 X_2})^b \PPE{u}{X_1} \PPE{v}{X_2}}
    =
    \includegraphicsbox[scale=1]{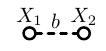}.
\end{equation}
We start by decomposing the level-one momentum generator
\begin{equation}
    \levo{P}^{\mu} = \sum^{n}_{j=1} \sum^{n}_{k=j+1} \levo{P}^{\mu}_{jk} - \sum^n_{j=1} s^{(1)}_j \levz{P}_j^{\mu} + \sum^n_{j=1} s^{(2)}_j \levz{P}_j^{\mu},
\end{equation}
where $s^{(1)}$ refers to the definition of the one-loop evaluation parameters as given in \eqref{eq:eval1loop}, but based on all external legs of the two-loop graph considered here. The bilocal term of the above operator can be further decomposed into
\begin{equation} \label{eq:Phattwoloopsplit}
    \sum^{n}_{j=1} \sum^{n}_{k=j+1} \levo{P}^{\mu}_{jk} = \sum_{j \in V_{X_1}} \sum_{\substack{k \in V_{X_1} \\ k > j}} \levo{P}^{\mu}_{jk} + \sum_{j \in V_{X_2}} \sum_{\substack{k \in V_{X_2} \\ k > j}} \levo{P}^{\mu}_{jk} + \levo{P}^{\mu}_{X_1 X_2}.
\end{equation}
Since the considered two-loop diagram is the simplest example of a graph with an end-vertex in the above sense, we can employ \eqref{eq:twopointsym} and \eqref{eq:endvertex} to get
\begin{equation}
    \levo{P}^{\mu} I_n^{(2)}
    = \sum_{j \in V_{X_1}} \left[ s^{(2)}_j - s^{(1)}_j \right] P^{\mu}_j I_n^{(2)} + \sum_{j \in V_{X_2}} \left[ s^{(2)}_j - s^{(1)}_j - \frac{1}{2} \left( D - 2b \right) \right] P^{\mu}_j I_n^{(2)}.
\end{equation}
Thus, defining the two-loop evaluation parameter according to
\begin{equation}
    s^{(2)}_j =
    \begin{cases}
        s^{(1)}_j & \text{for } j \in V_{X_1}, \\
        s^{(1)}_j + \frac{1}{2} \left( D - 2b \right) & \text{for } j \in V_{X_2},
    \end{cases}
\end{equation}
ensures level-one momentum symmetry for the two-loop Feynman diagram.

\paragraph{Three Loops.} 
Consider now an $n$-point three-loop Feynman integral in the form of
\begin{equation}
    I_n^{(3)} = \int \frac{d^Dx_{X_1} d^Dx_{X_2} d^Dx_{X_3}}{\left( x_{X_1 X_2}^2 \right)^{b_1} \left( x_{X_2 X_3}^2 \right)^{b_2} \PPE{u}{X_1} \PPE{v}{X_2} \PPE{w}{X_3}}
    =\includegraphicsbox[]{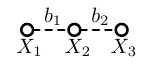}\,.
\end{equation}
When acting on this graph, the level-one momentum generator can be decomposed as
\begin{equation}
    \levo{P}^{\mu} = \sum^{n}_{j=1} \sum^{n}_{k=j+1} \levo{P}^{\mu}_{jk} - \sum^n_{j=1} s^{(1)}_j \levz{P}_j^{\mu} + \sum^n_{j=1} \Eval^{(3)}_{\text{track}, \, j} \levz{P}_j^{\mu},
\end{equation}
with the bilocal term being further divided into
\begin{equation}
    \sum^{n}_{j=1} \sum^{n}_{k=j+1} \levo{P}^{\mu}_{jk} = \sum^{3}_{p=1} \sum_{j \in V_{X_p}} \sum_{\substack{k \in V_{X_p} \\ k > j}} \levo{P}^{\mu}_{jk} + \levo{P}^{\mu}_{X_1 X_2} + \levo{P}^{\mu}_{X_1 X_3} + \levo{P}^{\mu}_{X_2 X_3}.
\end{equation}
The bilocal terms acting on the same vertex annihilate the integral due to \eqref{eq:twopointsym}, while one can use \eqref{eq:endvertex} and \eqref{eq:bridgevertex} to evaluate the cross-vertex bilocal terms:
\begin{equation}
    \begin{aligned}
        \levo{P}^{\mu}_{X_1 X_2} I_n^{(3)} &= -\frac{1}{2} \left( D - 2 b_1 \right) \levz{P}^{\mu}_{X_2} I_n^{(3)}, \\
        \levo{P}^{\mu}_{X_2 X_3} I_n^{(3)} &= \frac{1}{2} \left( D - 2 b_2 \right) \levz{P}^{\mu}_{X_2} I_n^{(3)}, \\
        \levo{P}^{\mu}_{X_1 X_3} I_n^{(3)} &= \frac{1}{2} \left( D - 2b_2 \right) \levz{P}^{\mu}_{X_1} I_n^{(3)} - \frac{1}{2} \left( D - 2b_1 \right) \levz{P}^{\mu}_{X_3} I_n^{(3)}.
    \end{aligned}
\end{equation}
Combining all contributions, one finds that the three-loop evaluation parameters necessary for level-one momentum symmetry of the graph, are given by
\begin{equation}
    s^{(3)}_{j} =
    \begin{cases}
        s^{(1)}_j - \frac{1}{2} \left( D - 2b_2 \right) & \text{for } j \in V_{X_1}, \\
        s^{(1)}_j + \frac{1}{2} \left( D - 2b_1 \right) - \frac{1}{2} \left( D - 2b_2 \right) & \text{for } j \in V_{X_2}, \\
        s^{(1)}_j + \frac{1}{2} \left( D - 2b_1 \right) & \text{for } j \in V_{X_3}.
    \end{cases}
\end{equation}
Now, due to level-zero translation invariance, one can shift the evaluation parameter by an overall constant which results in adding the level-zero generator,
\begin{equation}\label{eq:shift}
    \levo{P}^{\mu} \to \levo{P}^{\mu} + \kappa \levz{P}^{\mu},
\end{equation}
where $\kappa$ is a constant of choice. After setting $\kappa = \frac{1}{2} \left( D - 2b_2 \right)$, we finally obtain the convenient expressions
\begin{equation}
    s^{(3)}_{j} =
    \begin{cases}
        s^{(1)}_j & \text{for } j \in V_{X_1}, \\
        s^{(1)}_j + \frac{1}{2} \left( D - 2b_1 \right) & \text{for } j \in V_{X_2}, \\
        s^{(1)}_j + \frac{1}{2} \left( 2D - 2b_1 - 2b_2 \right) & \text{for } j \in V_{X_3}.
    \end{cases}
\end{equation}


\paragraph{General Decomposition of $\levo{P}$.}
The above examples suggest to introduce the following decomposition of the level-one momentum generator for a general $n$-point $\ell$-loop Feynman graph:
\begin{equation}\label{eq:levoP-decompose}
    \levo{P}^{\mu} = \sum^{\ell}_{p=1} \sum_{j \in V_{X_p}} \sum_{\substack{k \in V_{X_p} \\ k > j}} \levo{P}^{\mu}_{jk} + \sum^{\ell}_{p=1} \sum^{\ell}_{q=p+1} \levo{P}^{\mu}_{X_p X_q} - \sum^n_{j=1} s^{(1)}_j \levz{P}_j^{\mu} + \sum^n_{j=1} s^{(\ell)}_{j} \levz{P}_j^{\mu}.
\end{equation}
Here, the first term always annihilates the given integral using \eqref{eq:twopointsym} and the action of the second term must be evaluated using different bilocal symmetries.

\subsection{Stars}
With regard to the bilocal annihilators discussed in the previous section, a distinguished Feynman graph is an $n$-point vertex-star comprised of $\ell$ integration vertices with one central vertex and $\ell-1$ end-vertices:
\begin{equation}
    I^{(\ell)}_n = \includegraphicsbox{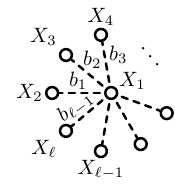}
\end{equation}
We thus have $\ell-1$ internal propagators with powers $b_1,\dots,b_{\ell-1}$ and each integration vertex may be connected to an arbitrary number of external legs.

By \eqref{eq:endvertex} and \eqref{eq:bridgevertex}, the action of the cross-vertex bilocal terms in \eqref{eq:levoP-decompose} yields
\begin{align}
    \levo{P}^{\mu}_{X_1 X_q} I^{(\ell)}_n &= \frac{1}{2} \left( D - 2 b_{q-1} \right) \levz{P}^{\mu}_{X_1} I^{(\ell)}_n, \\
    \levo{P}^{\mu}_{X_p X_q} I^{(\ell)}_{n} &= \frac{1}{2} \left( D - 2b_{q-1} \right) \levz{P}^{\mu}_{X_p} I^{(\ell)}_n - \frac{1}{2} \left( D - 2b_{p-1} \right) \levz{P}^{\mu}_{X_q} I^{(\ell)}_n,
\end{align}
where $1<p<q$. Combining them and shifting the evaluation parameters again via \eqref{eq:shift} with 
\begin{equation}
    \kappa = \frac{1}{2} \left( (\ell-1)D - 2 \sum^{\ell-1}_{q=1} b_q \right),
\end{equation}
we finally get
\begin{equation}\label{eq:star}
    s_{\text{star}, \, j}^{(\ell)} =
    \begin{cases}
        s_j^{(1)} & j \in V_{X_1}, \\
        s_j^{(1)} + \frac{1}{2} \left( D - 2 b_1 \right) & j \in V_{X_2}, \\
        s_j^{(1)} + \left( D - 2 b_1 \right) + \frac{1}{2} \left( D - 2 b_2 \right) & j \in V_{X_3}, \\
        \vdots & \vdots \\
        s_j^{(1)} + \left( (p-2)D - 2 \sum^{p-2}_{q=1} b_q \right) + \frac{1}{2} \left( D - 2 b_{p-1} \right) & j \in V_{X_p}, \\
        \vdots & \vdots \\
        s_j^{(1)} + \left( (\ell-3)D - 2 \sum^{\ell-3}_{q=1} b_q \right) + \frac{1}{2} \left( D - 2 b_{\ell-2} \right) & j \in V_{X_{\ell-1}}, \\
        s_j^{(1)} + \left( (\ell-2)D - 2 \sum^{\ell-2}_{q=1} b_q \right) + \frac{1}{2} \left( D - 2 b_{\ell-1} \right) & j \in V_{X_{\ell}},
    \end{cases}
\end{equation}
as the $\ell$-loop evaluation parameter required for level-one momentum symmetry of an $n$-point star with $\ell$ vertices.

\subsection{Tracks} 
Consider now an $n$-point Feynman diagram in the form of a track with $\ell$ integration vertices:
\begin{equation}
I^{(\ell)}_n 
=\includegraphicsbox[]{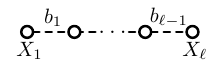}\;.
\end{equation}
These graphs generalize the family of train track integrals to tracks of arbitrary (not necessarily the same) integration vertices. In momentum space, these correspond to polygons glued to each other along a line.

Using \eqref{eq:endvertex-gen}, the cross-vertex bilocal terms in \eqref{eq:levoP-decompose} can be reduced as
\begin{equation}
    \sum^{\ell}_{q=p+1} \levo{P}^{\mu}_{X_p X_q} I^{(\ell)}_n = \frac{1}{2} \left( (\ell-p)D + 2 \sum^{\ell-1}_{q=p} b_{q} \right) \levz{P}^{\mu}_{X_p} I^{(\ell)}_n.
\end{equation}
After collecting the contributions for every $p$ and performing a shift \`a la \eqref{eq:shift} with 
\begin{equation}
    \kappa = \frac{1}{2} \left( (\ell-1)D - 2 \sum^{\ell-1}_{q=1} b_q \right),
\end{equation}
we finally get
\begin{equation}\label{eq:track}
    s_{\text{track}, \, j}^{(\ell)} =
    \begin{cases}
        s_j^{(1)} & j \in V_{X_1}, \\
        s_j^{(1)} + \frac{1}{2} \left( D - 2 b_1 \right) & j \in V_{X_2}, \\
        \vdots & \vdots \\
        s_j^{(1)} + \frac{1}{2} \left( (p-1)D - 2 \sum^{p-1}_{q=1} b_q \right) & j \in V_{X_p}, \\
        \vdots & \vdots \\
        s_j^{(1)} + \frac{1}{2} \left( (\ell-2)D - 2 \sum^{\ell-2}_{q=1} b_q \right) & j \in V_{X_{\ell-1}}, \\
        s_j^{(1)} + \frac{1}{2} \left( (\ell-1)D - 2 \sum^{\ell-1}_{q=1} b_q \right) & j \in V_{X_{\ell}},
    \end{cases}
\end{equation}
as the $\ell$-loop evaluation parameter necessary for level-one momentum symmetry of an $n$-point track with $\ell$ vertices.

\subsection{Track Networks (alias All Tree Graphs)}
\label{sec:alltrees}
We are now in the position to consider general tree graphs (black) corresponding to glued polygons in momentum space (green), e.g.\
\begin{equation}\label{fig:tree-example}
\includegraphicsbox[scale=1]{FigGeneralTreePolygonsMassless.pdf}
\quad\equiv\quad
\includegraphicsbox[scale=1]{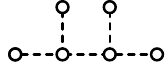}\;.
\end{equation}
Here, we once again illustrate our compactified notation on the right hand side.
We will show that also in the case of generic tree graphs, the above bilocal symmetries imply invariance under the full $\levo{P}$-operator with the prescription for the evaluation parameters as given in \secref{sec:Eval}.

\paragraph{General Tree.}
Consider a general $n$-point $\ell$-loop tree-like Feynman diagram in position space. Choosing any of the integration vertices to be the `root' vertex, the graph can be organized into `floors'\footnote{We do not use the term `levels' here in order to avoid confusion with the levels of the Yangian.} that are distinguished by their distance from the root, e.g.
\begin{equation}
    \includegraphicsbox{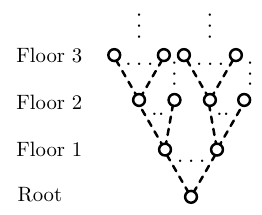}
\end{equation}
 A floor $F$ contains the vertices $\{ X^F_q \}$. Attached to a fixed vertex $X^F_p$ of a given floor is a set of sub-trees $\{ S^{F,p}_Q \}$ comprised of the attached vertices on higher floors.

Let $F$ be a floor (including the root), $X^F_p$ be a vertex of the floor and $S, S'$ be sub-trees attached to the vertex. Then, the differential operator
\begin{equation}
\label{eq:endvertexproof}
    \sum_{X \in S} \levo{P}^{\mu}_{X^F_p X},
\end{equation}
induces a generalized end-vertex symmetry, while the operator
\begin{equation}
\label{eq:bridgevertexproof}
    \sum_{X \in S} \sum_{Y \in S'} \levo{P}^{\mu}_{X Y},
\end{equation}
corresponds to a generalized bridge-vertex symmetry. Note that if the sub-trees are given by single vertices, then these generalized bilocal symmetries reduce to simple bilocal symmetries.

Now, we can prove the full level-one momentum symmetry of any given position-space tree by considering again the decomposition of the full level-one momentum generator as given in \eqref{eq:levoP-decompose}:
\begin{equation}
    \levo{P}^{\mu} = \sum^{\ell}_{p=1} \sum_{j \in V_{X_p}} \sum_{\substack{k \in V_{X_p} \\ k > j}} \levo{P}^{\mu}_{jk} + \sum^{\ell}_{p=1} \sum^{\ell}_{q=p+1} \levo{P}^{\mu}_{X_p X_q} - \sum^n_{j=1} \Eval^{(1)}_j \levz{P}_j^{\mu} + \sum^n_{j=1} \Eval^{(\ell)}_{j} \levz{P}_j^{\mu}.
\end{equation}
The goal is to reduce the bilocal terms in the first two summands to local terms that can be absorbed into the definition of the evaluation parameters $\Eval_j^{(\ell)}$. For the first bilocal term, this is trivial due to the two-point symmetries \eqref{eq:twopointsym}. It thus remains to consider the second bilocal term containing all ordered pairs of vertices $X_p^F$ and $X_q^{F'}$ on floors $F$ and $F'$. We will argue that every pair is contained exactly once in the set of operators given by \eqref{eq:endvertexproof} and \eqref{eq:bridgevertexproof}. There are two options: 
\begin{itemize}
    \item Either $X_q^{F'}\in S_P^{F,p}$, i.e.\ $X_q^{F'}$ belongs to the sub-tree attached to $X_p^{F}$, and the two vertices are connected by a straight line. In this case, the pair of vertices is contained in a generalized end-vertex symmetry \eqref{eq:endvertexproof} associated to $X_p^{F}$, e.g.\
    \begin{equation}
        \includegraphicsbox{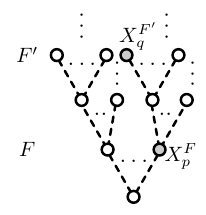}
    \end{equation}
    \item Or there exists a vertex $X_r^{F''}$ with $X_p^F\in S_R^{F'',r}$ and $X_q^{F'}\in S_{R'}^{F'',r}$ such that the pair $X_p^F$, $X_q^{F'}$ is connected by a line including a cusp at $X_r^{F''}$, and it is covered by a generalized bridge-vertex symmetry \eqref{eq:bridgevertexproof} associated to $X_r^{F''}$, e.g.\
    \begin{equation}
        \includegraphicsbox{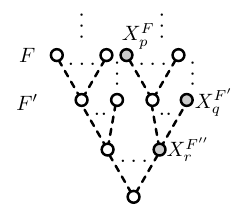}
    \end{equation}
\end{itemize}
Hence, invariance under the full level-one momentum generator $\levo{P}$ is traced back to the more elementary bilocal symmetries discussed above.

\paragraph{Evaluation Parameters.}
Let us now see that the above prescription leads to the recursive algorithm for the evaluation parameters given in \Secref{sec:Eval}.
To do so, consider an arbitrary vertex $X_p$ of a given tree graph. The line that connects this vertex with the root vertex $X_1$ divides the graph into three parts\footnote{Remember that we enumerate the vertices of a given graph by going clockwise around its boundary.}:
\begin{enumerate}[label=(\roman*)]
    \item vertices $X_q $ with $q<p$ (and attached propagators) that lie on the left of the line, 

    \item vertices $X_{q}$ with $q \leq p$ that lie on the line,

    \item vertices $X_q $ with $q>p$ that lie on the right of the line (including sub-trees attached to~$X_p$).
\end{enumerate}
Now, we want to collect all contributions from the above bilocal operators to the evaluation parameter associated with the considered vertex $X_p$. To get these contributions to $\Eval^{(\ell)}_{j}$ with $j \in V_{X_p}$, we must consider all bilocal symmetries, for which cross-vertex bilocal terms in \eqref{eq:levoP-decompose} reduce to local terms proportional to $\levz{P}_{X_p}^\mu$. For the vertices on the left of the line, one finds the following positive contributions to the evaluation parameters:
\begin{equation}
    \sum_{q<p} \levo{P}^{\mu}_{X_{q} X_{p}} I_n \propto -\frac{1}{2} \sum_{q<p} \left( D - 2 b_q \right) \levz{P}^{\mu}_{X_p} I_n \implies \Eval^{(\ell)}_{j} \to \Eval^{(\ell)}_{j} + \frac{1}{2} \sum_{q<p} \left( D - 2 b_q \right).
\end{equation}
The vertices that lie \emph{on} the line do not contribute because the resulting first order operators are not proportional to $\levz{P}_{X_p}^\mu$. For the vertices on the right of the line, one finds the following negative contributions:
\begin{equation}
    \sum_{q>p} \levo{P}^{\mu}_{X_{p} X_{q}} I_n \propto \frac{1}{2} \sum_{q>p} \left( D - 2 b_q \right) \levz{P}^{\mu}_{X_p} I_n \implies \Eval^{(\ell)}_{j} \to \Eval^{(\ell)}_{j} - \frac{1}{2} \sum_{q<p} \left( D - 2 b_q \right).
\end{equation}
Thus, we obtain the evaluation parameters $\Eval^{(\ell)}_{j}$ for  $j \in V_{X_p}$ as follows:
\begin{center}
\begin{minipage}{.9\textwidth}
    \emph{Start with $\Eval^{(\ell)}_{j} = \Eval^{(1)}_j$. 
    For every internal propagator with power $b$ on the left of the line, add $\frac{1}{2} \left( D - 2b \right)$.
    For every propagator on the right of the line, add $-\frac{1}{2} \left( D - 2b \right)$.} 
\end{minipage}
\end{center}
We then shift the resulting evaluation parameters by $\frac{1}{2} \left( D - 2b \right)$ for every internal propagator (with power $b$) in the entire graph i.e.
\begin{equation}
    \Eval^{(\ell)}_{j} \to \Eval^{(\ell)}_{j} + \frac{1}{2} \sum_{q} \left( D - 2 b_q \right).
\end{equation}
We can see that this shift eliminates all negative contributions from the right of the line, doubles all pre-existing positive contributions from the left of the line and adds positive contributions for the propagators lying on the line. This gives rise to an alternate (albeit more intuitive) prescription:
\begin{center}
\begin{minipage}{.9\textwidth}
    \emph{Start with $\Eval^{(\ell)}_{j} = \Eval^{(1)}_j$. 
    For every internal propagator with power $b$ on the left of the line, add $D - 2b$. 
    For every propagator on the line, add $\frac{1}{2} \left( D - 2b \right)$.}
\end{minipage} 
\end{center}
Finally, consider a pair of external legs $j\in V_{X_p}$ and $j+1\in V_{X_{p+1}}$.
To identify how the evaluation parameter changes when going from vertex $j$ to $j+1$, we calculate the difference $\Eval^{(\ell)}_{j+1} - \Eval^{(\ell)}_{j}$. Also here, we have two cases:
\begin{itemize}
    \item $X_p$ and $X_{p+1}$ are direct neighbors and are connected by a propagator with power~$b$. In this case, one can see that all (internal) propagators on the left of $X_p$ are also on the left of $X_{p+1}$. Moreover, $X_p$ lies on the line connecting $X_1$ and $X_{p+1}$ and the propagators on the right of $X_p$ include the one connecting $X_p$ and $X_{p+1}$, as well as all the propagators on the right of $X_{p+1}$. Thus, we have 
    \begin{equation}
       \Eval^{(\ell)}_{j+1} - \Eval^{(\ell)}_{j} = \Eval^{(1)}_{j+1} - \Eval^{(1)}_j - \left( -\frac{1}{2} \left( D - 2b \right) \right) = -\frac{a_j}{2} - \frac{a_{j+1}}{2} + \frac{1}{2} \left( D - 2b \right).
    \end{equation}

    \item $X_p$ and $X_{p+1}$ lie in different sub-trees of a common cusp vertex $X_r$. One now observes that the (internal) propagators on the left of $X_{p+1}$ are the ones between $X_r$ and $X_p$, plus all the propagators on the left of $X_p$. The propagators on the right of $X_p$ are the ones connecting $X_r$ and $X_{p+1}$, and all the propagators on the right of $X_{p+1}$. Thus, we get
    \begin{equation}
        \begin{aligned}
            \Eval^{(\ell)}_{j+1} - \Eval^{(\ell)}_{j} &= s^{(1)}_{j+1} - s^{(1)}_j + \frac{1}{2} \sum_{q \in \Gamma_1} \left( D - 2b_q \right) - \left( -\frac{1}{2} \sum_{q \in \Gamma_2} \left( D - 2b_q \right) \right)
            \\
            &= -\frac{a_j}{2} - \frac{a_{j+1}}{2} + \frac{1}{2} \sum_{q \in \Gamma} \left( D - 2b_q \right),
    \end{aligned}
    \end{equation}
    where $\Gamma_1$ denotes the path between $X_r$ and $X_p$, $\Gamma_2$ the path between $X_r$ and $X_{p+1}$ and $\Gamma = \Gamma_1 \cup \Gamma_2$ the path between $X_p$ and $X_{p+1}$. 
\end{itemize}
We therefore conclude that the prescription for obtaining the evaluation parameters as given in \secref{sec:Eval} follows from the elementary bilocal symmetries contributing to the full level-one momentum invariance of the graph.

\paragraph{Example.}
Let us illustrate the above by considering the example \eqref{fig:tree-example} of an $n$-point $6$-loop diagram, which can be drawn as 
\begin{equation}
    \includegraphicsbox{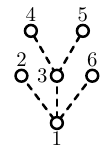}
\end{equation}
Following the above logic, we identify all bilocal contributions to the full level-one momentum, which induce symmetries of the graph on their own. 

For the \emph{Root}, we have
    \begin{itemize}
        \item Simple end-vertex symmetries: $X_1, S = \{ X_2 \}$ and $X_1, S = \{ X_6 \}$.
        \begin{equation}
            \begin{aligned}
                \levo{P}^{\mu}_{X_1 X_2} I^{(6)}_n &= \frac{1}{2} \left( D - 2b_{1,2} \right) \levz{P}^{\mu}_{X_1} I^{(6)}_n, \\
                \levo{P}^{\mu}_{X_1 X_6} I^{(6)}_n &= \frac{1}{2} \left( D - 2b_{1,6} \right) \levz{P}^{\mu}_{X_1} I^{(6)}_n.
            \end{aligned}
        \end{equation}
        \item Generalized end-vertex symmetry: $X_1, S = \{ X_3, X_4, X_5 \}$.
        \begin{equation}
            \sum^{5}_{q=3} \levo{P}^{\mu}_{X_1 X_q} I^{(6)}_n = \frac{1}{2} \left( 3D - 2b_{1,3} - 2b_{3,4} - 2b_{3,5} \right) \levz{P}^{\mu}_{X_1} I^{(6)}_n.
        \end{equation}

        \item Simple bridge-vertex symmetry: $X_1, S = \{ X_2 \}, S' = \{ X_6 \}$.
        \begin{equation}
            \levo{P}^{\mu}_{X_2 X_6} I^{(6)}_n = \frac{1}{2} \left( D - 2b_{1,6} \right) \levz{P}^{\mu}_{X_2} I^{(6)}_n - \frac{1}{2} \left( D - 2b_{1,2} \right) \levz{P}^{\mu}_{X_6} I^{(6)}_n.
        \end{equation}

        \item Generalized bridge-vertex symmetries: $X_1, S = \{ X_2 \}, S' = \{ X_3, X_4, X_5 \}$ and $X_1, S = \{ X_3, X_4, X_5 \}, S' = \{ X_6 \}$.
        \begin{equation}
            \begin{aligned}
                \sum^{5}_{q=3} \levo{P}^{\mu}_{X_2 X_q} I^{(6)}_n &= \frac{1}{2} \left( 3D - 2b_{1,3} - 2b_{3,4} - 2b_{3,5} \right) \levz{P}^{\mu}_{X_2} I^{(6)}_n - \frac{1}{2} \left( D - 2b_{1,2} \right) \sum^{5}_{q=3} \levz{P}^{\mu}_{X_q} I^{(6)}_n, \\
                \sum^{5}_{q=3} \levo{P}^{\mu}_{X_q X_6} I^{(6)}_n &= \frac{1}{2} \left( D - 2b_{1,6} \right) \sum^{5}_{p=3} \levz{P}^{\mu}_{X_p} I^{(6)}_n - \frac{1}{2} \left( 3D - 2b_{1,3} - 2b_{3,4} - 2b_{3,5} \right) \levz{P}^{\mu}_{X_6} I^{(6)}_n.
            \end{aligned}
        \end{equation}
        \end{itemize}
        
For \emph{Floor~1}, we find
    \begin{itemize}
        \item Simple end-vertex symmetries: $X_3, S = \{ X_4 \}$ and $X_3, S = \{ X_5 \}$.
        \begin{equation}
            \begin{aligned}
                \levo{P}^{\mu}_{X_3 X_4} I^{(6)}_n &= \frac{1}{2} \left( D - 2b_{3,4} \right) \levz{P}^{\mu}_{X_3} I^{(6)}_n, \\
                \levo{P}^{\mu}_{X_3 X_5} I^{(6)}_n &= \frac{1}{2} \left( D - 2b_{3,5} \right) \levz{P}^{\mu}_{X_3} I^{(6)}_n.
            \end{aligned}
        \end{equation}

        \item Simple bridge-vertex symmetry: $X_3, S = \{ X_4 \}, S' = \{ X_5 \}$.
        \begin{equation}
            \levo{P}^{\mu}_{X_4 X_5} I^{(6)}_n = \frac{1}{2} \left( D - 2b_{3,5} \right) \levz{P}^{\mu}_{X_4} I^{(6)}_n - \frac{1}{2} \left( D - 2b_{3,4} \right) \levz{P}^{\mu}_{X_5} I^{(6)}_n.
        \end{equation}
    \end{itemize}
Combining the action of these operators and shifting the result via \eqref{eq:shift} with
\begin{equation}
    \kappa = \frac{1}{2} \left( 5D - 2b_{1,2} - 2b_{1,3} - 2b_{3,4} - 2b_{3,5} - 2b_{1,6} \right),
\end{equation}
allows us to build the full level-one momentum generator with the explicit evaluation parameters given by
\begin{equation}
    s^{(6)}_{j} =
    \begin{cases}
        s^{(1)}_j & \text{for } j \in V_{X_1}, \\
        s^{(1)}_j + \frac{1}{2} \left( D - 2b_{1,2} \right) & \text{for } j \in V_{X_2}, \\
        s^{(1)}_j + \frac{1}{2} \left( 3D - 4b_{1,2} - 2b_{1,3} \right) & \text{for } j \in V_{X_3}, \\
        s^{(1)}_j + \frac{1}{2} \left( 4D - 4b_{1,2} - 2b_{1,3} - 2b_{3,4} \right) & \text{for } j \in V_{X_4}, \\
        s^{(1)}_j + \frac{1}{2} \left( 6D - 4b_{1,2} - 2b_{1,3} - 4b_{3,4} - 2b_{3,5} \right) & \text{for } j \in V_{X_5}, \\
        s^{(1)}_j + \frac{1}{2} \left( 9D - 4b_{1,2} - 4b_{1,3} - 4b_{3,4} - 4b_{3,5} - 2b_{1,6} \right) & \text{for } j \in V_{X_6}.
    \end{cases}
\end{equation}
This is line with the prescription given in \secref{sec:Eval}.

\section{Momentum Space Interpretation of Full $\levo{P}$-Symmetry}\label{sec:MomSpace}
So far our calculations were performed in position (or region momentum) space. In this section, we review (see \cite{Loebbert:2020hxk}) the translation of the above full level-one symmetry discussed in the previous \Secref{sec:full} to momentum space ($p$-space) with the momenta defined via 
\begin{equation}\label{eq:dualps}
    p_j = x_j - x_{j+1}.
\end{equation}
For an $n$-point Feynman graph, we furthermore assume the boundary condition 
\begin{equation}
    x^{\mu}_{n+1} \equiv x^{\mu}_{1}.
\end{equation}
In the following, we will demonstrate that the momentum space conformal algebra arises under this map from the $x$-space generators according to the rules\footnote{Note that these generators essentially coincide with the Fourier transform of the $x$-space generators given in \eqref{eq:levzdensities}.} 
\begin{equation}\label{eq:levzinpspace}
    \begin{aligned}
        \levz{P}^\mu &\to \bar{\levz{P}}^{\mu} = \sum_j p_j^{\mu}, \\
        \levz{L}^{\mu \nu} &\to \bar{\levz{L}}^{\mu \nu} = \sum_j \left( p^{\mu}_j \bar{\partial}^{\nu}_j - p^{\nu}_j \bar{\partial}^{\mu}_j \right), \\
        \levz{D} &\to \bar{\levz{D}} = \sum_j \left( p_{j \mu} \bar{\partial}^{\mu}_j + \bar{\Delta}_j \right), \\
        \levo{P}^\mu &\to \bar{\levz{K}}^{\mu} = \sum_j \left( p_j^{\mu} \bar{\partial}^2_j - 2 p_{j \nu} \bar{\partial}^{\nu}_j \bar{\partial}^{\mu}_j - 2 \bar{\Delta}_j \bar{\partial}^{\mu}_j \right),
    \end{aligned}
\end{equation}
with constants $\bar{\Delta}_j$ to be specified below and the $\levo{P}$ identification applies when acting on the integral. To see this, we invert the above definition of the momenta \eqref{eq:dualps} to find
\begin{equation}
    x_i^{\mu} = x_1^{\mu} - \sum_{j=1}^{i-1} p^{\mu}_j,
\end{equation}
such that the boundary condition turns into momentum conservation
\begin{equation}\label{eq:p-conserve}
    \sum_{i=1}^n p^{\mu}_i
    = \sum_{i=1}^n \left( x^{\mu}_i - x^{\mu}_{i+1} \right)
    = x^{\mu}_1 - x^{\mu}_1
    = 0.
\end{equation}
Similarly, we transform partial derivatives according to
\begin{equation}
    \partial^{\mu}_i
    = \sum_{j=1}^n \frac{\partial p_{j \nu}}{\partial x_{i \mu}} \frac{\partial}{\partial p_{j \nu}}
    = \sum_{j=1}^n \left( \delta_{i,j} - \delta_{i,j+1} \right) \delta^{\mu}_{\nu} \frac{\partial}{\partial p_{j \nu}}
    = \bar{\partial}^{\mu}_{i} - \bar{\partial}^{\mu}_{i-1},
    \label{eq:derivtrans}
\end{equation}
where $\bar{\partial}^{\mu}_i := \frac{\partial}{\partial p_{i \mu}}$. Using the above transformations and \eqref{eq:doublecounting}, one can express the (level-zero) momentum, Lorentz and dilation generator as
\begin{equation}
    \begin{aligned}
        i\levz{P}^{\mu} &\to \bar{\partial}^{\mu}_{n} - \bar{\partial}^{\mu}_{0} = 0 = \sum_{j=1}^n p^{\mu}_{j} = \bar{\levz{P}}^{\mu}, \\
        -i\levz{L}^{\mu \nu} &\to \sum^n_{j=1} \left( p^{\mu}_j \bar{\partial}^{\nu}_j - p^{\nu}_j \bar{\partial}^{\mu}_j \right) = \bar{\levz{L}}^{\mu \nu}, \\
        i\levz{D} &\to \sum^n_{j=1} \left( p_{j \mu} \bar{\partial}^{\mu}_j + \Delta_j \right) = \bar{\levz{D}}_{\bar{\Delta} = 0} + \sum^n_{j=1} \Delta_j,
    \end{aligned}
\end{equation}
where $\bar{\levz{D}}_{\bar{\Delta} = 0} = \sum^n_{j=1} p_{j \mu} \bar{\partial}^{\mu}_j$. Thus, under the transformation \eqref{eq:dualps} the generators are mapped according to \eqref{eq:levzinpspace}, with the identification
\begin{equation}
    \sum^n_{j=1} \bar{\Delta}_j := \sum^n_{j=1} \Delta_j.
\end{equation}
Let us now consider the level-one momentum generator with its first- and second-order (derivative) contributions given by
\begin{equation}
    \begin{aligned}
        \levo{P}^{\mu}_{\text{first-order}} &= \frac{1}{2} \sum^n_{j=1} \sum^n_{k=j+1} \left( \Delta_k \levz{P}_j^{\mu} - \Delta_j \levz{P}_k^{\mu} \right) + \sum_{j=1}^n s_j \levz{P}_j^{\mu}, \\
        \levo{P}^{\mu}_{\text{second-order}} &= \frac{i}{2} \sum^n_{j=1} \sum^n_{k=j+1} \left( \levz{P}_{j \nu} \left( \eta^{\mu \nu} \levz{D}_{k, \, \Delta = 0} + \levz{L}^{\mu \nu}_k \right) - \left( j \leftrightarrow k \right) \right),
    \end{aligned}
\end{equation}
where $\levz{D}_{k, \, \Delta = 0} = -i x_{k \nu} \partial^{\nu}_k$. Transforming the above operators, we find
\begin{equation}
    \begin{aligned}
        -2i \, \levo{P}^{\mu}_{\text{first-order}} &\to -\sum_{j=1}^n \left( \Delta_j + \Delta_{j+1} + 2 s_j - 2 s_{j+1} \right) \bar{\partial}_j^{\mu} + 2 \sum_{j=1}^n \Delta_j \bar{\partial}_n^{\mu}, \\
        -2i \, \levo{P}^{\mu}_{\text{second-order}} &\to \sum_{j=1}^n \left( p_j^{\mu} \bar{\partial}^2_j - 2 p_{j \nu} \bar{\partial}^{\nu}_j \bar{\partial}^{\mu}_j \right) + 2 \sum_{j=1}^n \left( \eta^{\mu \nu} \bar{\levz{D}}_{j, \, \bar{\Delta} = 0} - \bar{\levz{L}}^{\mu \nu}_j \right) \bar{\partial}_{n \nu},
    \end{aligned}
\end{equation}
where we make the identifications $\Delta_{n+1} \equiv \Delta_1$ and $s_{n+1} \equiv s_1$. Adding up the different contributions, we get
\begin{equation}
    \begin{aligned}
    -2i \levo{P}^{\mu} \to& \sum^n_{j=1} \left[ \bar{\levz{K}}^{\mu}_{j, \, \bar{\Delta} = 0} - \left( \Delta_j + \Delta_{j+1} + 2 s_j - 2 s_{j+1} \right) \bar{\partial}_j^{\mu} \right]
    \\
    &+ 2 \sum^n_{j=1} \left[ \eta^{\mu \nu} \left( \bar{\levz{D}}_{j}^{\bar{\Delta} = 0} + \Delta_j \right) - \bar{\levz{L}}^{\mu \nu}_j \right] \bar{\partial}_{n \nu},
\end{aligned}
\end{equation}
where $\bar{\levz{K}}^{\mu}_{j, \, \bar{\Delta} = 0} = p_j^{\mu} \bar{\partial}^2_j - 2 p_{j \nu} \bar{\partial}^{\nu}_j \bar{\partial}^{\mu}_j$. Now, consider the transformation of a general $n$-point Feynman integral in $x$-space to the dual $p$-space
\begin{equation}
    I_n(x_1, \dots, x_n) \to I_n(p_1, \dots, p_{n-1}),
\end{equation}
where the $p_n$-dependence of $I_n$ can be eliminated using \eqref{eq:p-conserve}. This implies
\begin{equation}
    \levo{P}^{\mu} I_n(x_1, \dots, x_n) \to \bar{\levz{K}}^{\mu} I_n(p_1, \dots, p_{n-1}),
\end{equation}
with the identification \cite{Loebbert:2020hxk}
\begin{equation}
    \bar{\Delta}_i \equiv \frac{1}{2} \left( \Delta_i + \Delta_{i+1} + 2 s_i - 2 s_{i+1} \right).
\end{equation}
Performing a consistency check, we find
\begin{equation}
    \sum^n_{i=1} \bar{\Delta}_i = \frac{1}{2} \sum^n_{i=1} \left( \Delta_i + \Delta_{i+1} + 2 s_i - 2 s_{i+1} \right) = \sum^n_{i=1} a_i = \sum^n_{i=1} \Delta_i.
\end{equation}
Here, we employ the evaluation parameters as given in \Secref{sec:Eval}.

\section{Massive Generalization}
In this section, we discuss how the symmetries we have obtained so far extend to diagrams with massive propagators on the boundaries of the considered Feynman graphs.

The massive $x$-space representation of the conformal generator densities corresponds to the $(D+1)$-dimensional conformal algebra and is given by \cite{Alday:2009zm}
\begin{equation}\label{eq:levzdensities+mass}
    \begin{aligned}
    \levz{P}^{\mu}_j &= -i \partial^{\mu}_j, \\
    \levz{L}^{\mu \nu}_j &= i( x^{\mu}_j \partial^{\nu}_j - x^{\nu}_j \partial^{\mu}_j ), \\
    \levz{D}_j &= -i( x_{j \mu} \partial^{\mu}_j + m_j \partial_{m_j} + \Delta_j ), \\
    \levz{K}^{\mu}_j &= -i(2 x^{\mu}_j (x^{\nu}_j \partial_{j \nu} + m_j \partial_{m_j}) - (x^2_j + m^2_j) \partial^{\mu}_j + 2 \Delta_j x^{\mu}_j),
    \end{aligned}
\end{equation}
where $\displaystyle \partial_{m_j} := {\partial}/{\partial m_j}$ with $m_j$ being the mass of the $j$-th external leg.
The level-zero invariance under the massive generators can be shown in complete analogy to the massless case, if we replace the propagators in the Feynman integrals by massive propagators, and set masses associated with internal vertices to zero, cf.\ \cite{Loebbert:2020hxk,Loebbert:2020glj}:
\begin{equation}
    \frac{1}{x_{jk}^2} \to \frac{1}{x_{jk}^2 + (m_j-m_k)^2}, \qquad  m_\text{internal} = 0.
\end{equation}
For example, a massive $n$-point one-loop integral takes the form
\begin{equation}
    I_n^{m_j} = \int \frac{d^D x_0}{\prod_{j=1}^n (x_{j0}^2 + m_j^2)^{a_j}}.
\end{equation}

In general, substituting the massive level-zero generator densities into the bilocal level-one momentum generator density \eqref{eq:levobilocal} gives
\begin{equation}
    \levo{P}^{\mu}_{jk} = \frac{i}{2} \left( x_{jk \nu} \partial_j^{\mu} \partial_k^{\nu} - x_{jk}^{\mu} \partial_j^{\nu} \partial_{k \nu} + x_{jk \nu} \partial_{j}^{\nu} \partial_k^{\mu} - m_k \partial_{m_k} \partial_j^{\mu} + m_j \partial_{m_j} \partial_k^{\mu} - 2 a_k \partial_j^{\mu} + 2 a_j \partial_k^{\mu} \right).
\end{equation}
The action of this operator on the product of two massive propagators yields
\begin{equation}
    \levo{P}^{\mu}_{jk} \left( \frac{1}{(x^2_{j X} + m_j^2)^{a_j} (x^2_{k Y} + m_k^2)^{a_k}} \right) = 2 i a_j a_k \frac{T^{\alpha \beta \mu \nu} x_{X Y \nu} \ x_{j X \alpha} \ x_{k Y \beta}}{(x^2_{j X} + m_j^2)^{a_j + 1} (x^2_{k Y} + m_k^2)^{a_k + 1}},
    \label{eq:Phattwoprop+m}
\end{equation}
in complete analogy to the massless case. Thus, all bilocal symmetries discussed in \Secref{sec:Bilocal}, as well as the full level-one momentum symmetry, extend to the massive case.

We remark that in the massive case, the employed expression \eqref{eq:levoP} is not the full $(D+1)$-dimensional level-one momentum generator, since the contraction of momentum and Lorentz generator is only performed over $D$ dimensions. In fact, the full $(D+1)$-dimensional level-one momentum reads
\begin{equation}\label{eq:levoP+mass}
    \levo{P}^{\mu} = \frac{i}{2} \sum^n_{j=1} \sum^n_{k=j+1} \left( \levz{P}^{\mu}_j \levz{D}_k + \levz{P}_{j \nu} \levz{L}^{\mu \nu}_k - \levz{D}_j \levz{P}^{\mu}_k - \levz{L}^{\mu \nu}_j \levz{P}_{k \nu} \right) + \sum^n_{j=1} \Eval_j \levz{P}_j^{\mu} + \levo{P}^{\mu}_{\text{extra}},
\end{equation}
with the `extra' operator being defined as
\begin{equation}
    \levo{P}^{\mu}_{\text{extra}} = \frac{i}{2} \sum^n_{j=1} \sum^n_{k=j+1} \left( \levz{P}_{j D+1} \levz{L}^{\mu D+1}_k - \levz{L}^{\mu D+1}_j \levz{P}_{k D+1} \right),
\end{equation}
where
\begin{equation}
    \levz{P}_{j D+1} = -i \partial_{m_j}, \qquad \levz{L}^{\mu D+1}_j = i( x^{\mu}_j \partial_{m_j} - m_j \partial^{\mu}_j ).
\end{equation}
For a more detailed discussion of $\levo{P}^{\mu}_{\text{extra}}$, see \cite{Loebbert:2020glj}.

\section{Conclusions and Outlook}
In this paper, we have shown that tree-level Feynman graphs in position space (alias multi-loop integrals in dual momentum space) are annihilated by a number of bilocal differential operators of Yangian type. All of these correspond to second order partial differential equations that constrain the considered Feynman integral. In particular, these can be combined to obtain the full level-one momentum invariance, with the Yangian generator $\levo{P}^\mu$ acting on all external legs of a given graph. For the full level-one momentum symmetry, we have also reviewed an alternative interpretation in terms of a momentum space conformal symmetry. We are not aware of such an interpretation for the partial bilocal symmetries discussed in \Secref{sec:Bilocal}. All considered symmetries hold for graphs with external propagators massive or massless, and massless internal propagators. These findings suggest a number of future research directions.
\medskip

\begin{figure}[t] 
\begin{center}
\includegraphicsbox{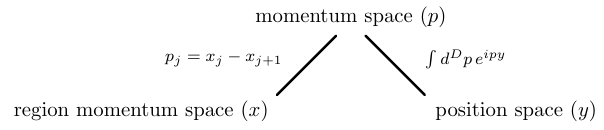}
\end{center}
\caption{The symmetries discussed in this paper can be considered in different spaces. Here, the duality transformation on the left edge may turn local symmetries into non-local symmetries or vice versa (cf.\ \protect\Secref{sec:MomSpace}), while the Fourier transform on the right edge may turn first order differential operators into second order differential operators, e.g.\ the special conformal generator.}
\label{fig:threespaces}
\end{figure}

The interpretation of the level-one momentum generator $\levo{P}^\mu$ as the special conformal generator in momentum space suggests to also investigate the Fourier-transform of the considered integrals in momentum-space (as opposed to the dual $x$-space related via $p_j=x_j-x_{j+1}$), see \Figref{fig:threespaces}. E.g.\ for three points and one loop, the momentum space Feynman integral Fourier transforms into the well known expression for a three-point function with position space (here, $y$-space) conformal symmetry~\cite{Coriano:2013jba}:
\begin{equation}
I_3(x) 
\quad
\begin{matrix}
    \scriptstyle\text{dual}
    \\
\longrightarrow
    \\
    \scriptstyle{p_j=x_j-x_{j+1}}
\end{matrix}\quad
I_3(p) 
\quad
\begin{matrix}
    \scriptstyle\text{Fourier}
    \\
\longrightarrow
    \\
    \scriptstyle{\int d^Dp\, e^{ipy}}
\end{matrix}\quad
\frac{c_{123}}{y_{12}^{2A}y_{13}^{2B}y_{23}^{2C}}.
\end{equation}
Here, $y$ denotes the Fourier transformed $p$-variables, while $x$ represents the region momenta defined via $p_j=x_j-x_{j+1}$.
Hence, we can understand the momentum space conformal symmetry of this one-loop integral as the Fourier transform of the $y$-space conformal symmetry. 
In this context, it should be noted that the momentum space conformal constraints were solved in terms of simplex integrals in \cite{Bzowski:2019kwd,Bzowski:2020kfw}. It would be interesting to explicitly map these simplices to the considered Feynman integrals and to extend them to the situation with massive propagators discussed in the present paper. In a similar direction, it would be useful to connect the above discussions to the recent position space analysis of Ward identities for Feynman integrals based on insights on the so-called banana graphs~\cite{Mishnyakov:2023wpd,Mishnyakov:2024rmb,Mishnyakov:2024xjz}.
Also, the interesting relation to the Gelfand--Kapranov--Zelevinsky (GKZ) approach towards Feynman integrals should be further explored, which in the context of the Yangian was recently presented in \cite{Talk:Mishnyakov}, see also similar structures discussed in the context of contact Witten diagrams \cite{Rigatos:2022eos}.
\medskip

It is expected that under certain conditions also loop diagrams in position space are invariant under the above (partial) bilocal operators or the full momentum space conformal symmetry. On the one hand, this is motivated by the fact that certain position-space loop diagrams form part of the family of fishnet graphs, which were shown to be Yangian invariant for specific propagator powers \cite{Chicherin:2017cns,Chicherin:2017frs,Kazakov:2023nyu}.  On the other hand, this is also suggested by the above relation to (Fourier-transformed) momentum space correlators and their simplex interpretation \cite{Bzowski:2019kwd,Bzowski:2020kfw}. 
It will be interesting to further analyze this point in full detail. 
On that note, it would also be useful to generalize the Yangian proofs based on the lasso method \cite{Chicherin:2017cns,Chicherin:2017frs,Kazakov:2023nyu}, as well as the considerations of momentum space conformal symmetry, to the case including massive propagators on the boundary of the Feynman graph.
\medskip

The considered position space Feynman graphs typically feature a number of permutation symmetries $\sigma$ in the external legs $x_j$, which do not commute with the level-one momentum generator. Application of these permutation symmetries to the level-one momentum $\levo{P}$ thus results in new differential operators $\sigma\levo{P}\sigma^{-1}$ that annihilate the integral. 
In \cite{Duhr:2022pch,Duhr:2024hjf}, it was shown that this approach leads to the full Picard--Fuchs ideal defining the Calabi--Yau geometry underlying certain families of fishnet integrals. It would be helpful to understand in how far the bilocal annihilators found in the present paper relate to the combination of full $\levo{P}$-symmetry combined with permutations, and whether the two sets of differential operators can be matched.
\medskip

The families of multi-loop Feynman integrals considered in this paper are highly complicated and only the simplest examples have been computed so far. It is thus natural to ask in how far the symmetries discussed here can help to gain better control over these integrals. It has been shown on several examples that Yangian symmetry allows to bootstrap Feynman integrals from scratch 
\cite{
Loebbert:2019vcj, Loebbert:2020hxk, Loebbert:2020glj, Loebbert:2020aos, Duhr:2022pch, Duhr:2023eld, Duhr:2024hjf}. The philosophy of this Yangian bootstrap is that the level-zero or dual conformal symmetry constrains the independent variables to the conformal cross ratios. Then, the level-one Yangian symmetry implies second order differential equations for the resulting functions of these conformal variables. Dropping the level-zero symmetry (as done in the present paper) thus implies that the integral depends on a larger set of non-conformal variables, but the level-one symmetry still yields constraining differential equations. It is thus natural to ask in how far these differential equations constrain the integral, and whether there are similar relations to (Calabi--Yau) geometry as discovered in the case with a full Yangian symmetry, see 
\cite{Duhr:2022pch,Duhr:2023eld,Duhr:2024hjf}. In fact, some examples of one-loop integrals without (level-zero) conformal symmetry have been bootstrapped in this way, cf.\ \cite{Loebbert:2020glj,Loebbert:2020aos}. 
\medskip

Finally, it is worth mentioning that the above bootstrap approach for Feynman integrals resembles the way conformal partial waves are defined via Casimir operators in the context of conformal field theory, see e.g.\ \cite{Dolan:2011dv}. While conformal partial waves are defined as eigenstates of Casimir operators, here we employ a different algebraic structure, which originates in the Yangian, in order to constrain Feynman integrals:
\begin{equation}
  \stackrel{\text{Casimirs}}{\levz{J}_{j,a}\levz{J}^a_{k}} \qquad\longleftrightarrow\qquad \stackrel{\text{Yangian bilocals}}{f^a{}_{bc}\,\levz{J}^c_j \levz{J}^b_k}.
\end{equation}
The present paper suggests to further explore this analogy and to think of Feynman integrals as being defined via the bilocal differential annihilators found above. At least in two spacetime dimensions, the results of \cite{Duhr:2022pch,Duhr:2024hjf} show that these differential operators completely constrain full families of multi-loop Feynman integrals.

\subsection*{Acknowledgements} 
We are grateful to Claude Duhr and Sven Stawinski for various related discussions. Moreover, we thank Claude Duhr, Gwena\"el Ferrando and Sven Stawinski for helpful comments on the manuscript.

\appendix

\section{Conformal Commutation Relations}\label{app:confalg}
The position space conformal generators employed in this paper satisfy the following commutation relations:
\begin{align}
   \comm{ \levz{P}^{\mu}}{ \levz{L}^{\nu \rho} } &= i (\eta^{\mu \nu} \levz{P}^{\rho} - \eta^{\mu \rho} \levz{P}^{\nu}),
   \\
    \comm{\levz{L}^{\mu \nu}}{\levz{L}^{\rho \sigma}} &= i (\eta^{\nu \rho} \levz{L}^{\mu \sigma} + \eta^{\mu \sigma} \levz{L}^{\nu \rho} - \eta^{\mu \rho} \levz{L}^{\nu \sigma} - \eta^{\nu \sigma} \levz{L}^{\mu \rho}),
    \\
    \comm{\levz{K}^{\mu}}{\levz{P}^{\nu}} &= 2i (\eta^{\mu \nu} D - \levz{L}^{\mu \nu}),
    \\
    \comm{\levz{K}^{\mu}}{\levz{L}^{\nu \rho}} &= i (\eta^{\mu \nu} \levz{K}^{\rho} - \eta^{\mu \rho} \levz{K}^{\nu}),
    \\
    \comm{\levz{D}}{\levz{P}^{\mu}} &= +i \levz{P}^{\mu}, 
    \\
    \comm{\levz{D}}{\levz{K}^{\mu}} &= -i \levz{K}^{\mu},
\end{align}
with the rest of the commutators vanishing.

\section{Details on the Yangian}\label{app:lev1densities}
The two-point level-one Yangian generator densities are given by
\begin{align}
  \levo{P}^{\mu}_{jk} &= \frac{i}{2} \left( \levz{P}^{\mu}_j \levz{D}_k + \levz{P}_{j \nu} \levz{L}^{\mu \nu}_k - i a_k \levz{P}^{\mu}_j - (j \leftrightarrow k) \right), \\
  \levo{L}^{\mu \nu}_{jk} &= \frac{i}{2} \left( \frac{1}{2} (\levz{P}^{\mu}_j \levz{K}^{\nu}_k - \levz{P}^{\nu}_j \levz{K}^{\mu}_k) + \levz{L}^{\mu \rho}_k {\levz{L}_{j \rho}}^{\nu} - i a_k \levz{L}^{\mu \nu}_j - (j \leftrightarrow k) \right), \\
  \levo{D}_{jk} &= \frac{i}{2} \left( \frac{1}{2} \levz{P}_{j \mu} \levz{K}^{\mu}_k - i a_k \levz{D}_j - (j \leftrightarrow k) \right), \\
  \levo{K}^{\mu}_{jk} &= \frac{i}{2} \left( \levz{K}^{\mu}_k \levz{D}_j + \levz{K}_{j \nu} \levz{L}^{\mu \nu}_k - i a_k \levz{K}^{\mu}_j - (j \leftrightarrow k) \right).
\end{align}
The Serre relations take the form
\begin{align}
    [ \levo{J}^a, [ \levo{J}^b, \levz{J}^c ]] - [ \levz{J}^a, [ \levo{J}^b, \levo{J}^c ]] &= {g^{abc}}_{def} \{ \levz{J}^d, \levz{J}^e, \levz{J}^f \}, \\
    [[ \levo{J}^a, \levo{J}^b ], [ \levz{J}^r, \levo{J}^s ]] + [[ \levo{J}^r, \levo{J}^s ], [ \levz{J}^a, \levo{J}^b ]] &= ({g^{abc}}_{def} {f^{rs}}_c + {g^{rsc}}_{def} {f^{ab}}_c) \{ \levz{J}^d, \levz{J}^e, \levz{J}^f \},
\end{align}
where
\begin{equation}
    {g^{abc}}_{def} = \frac{1}{24} {f^a}_{di} {f^b}_{ej} {f^c}_{fk} f^{ijk}, \qquad \{ \levz{X}_1, \levz{X}_2, \levz{X}_3 \} = \sum_{i \neq j \neq k = 1}^{3} \levz{X}_i \levz{X}_j \levz{X}_k.
\end{equation}
For the Serre relations and the conformal representation discussed here, see \cite{Miczajka:2022jjv,Dokmetzoglou:2022mfd}.

\bibliographystyle{nb}
\bibliography{YangianBilocals}

\end{document}